\documentclass[12pt]{article}

\newenvironment{wileykeywords}{\textsf{Keywords:}\hspace{\stretch{1}}}{\hspace{\stretch{1}}\rule{1ex}{1ex}}

\usepackage{epsfig,multirow}
\usepackage{tikz}
\usepackage{amsmath,amssymb}
\usepackage{graphicx}
\usepackage{color}
\usepackage{dcolumn}
\usepackage{bm}
\usepackage[numbers,super,comma,sort&compress]{natbib}

\definecolor{background-color}{gray}{0.98}

\title{Comparison of the MSMS and NanoShaper molecular surface triangulation codes in 
the TABI Poisson--Boltzmann solver}
\author{Leighton Wilson\thanks{Department of Mathematics, University of Michigan, Ann Arbor, MI 48109}, Robert Krasny\thanks{Department of Mathematics, University of Michigan, Ann Arbor, MI 48109}}

\begin{document}

\maketitle

\begin{abstract}
The Poisson-Boltzmann (PB) implicit solvent model is a popular framework for studying
the electrostatics of biomolecules immersed in water with dissolved salt.
In this model the dielectric interface between the biomolecule and solvent
is often taken to be the molecular surface or solvent-excluded surface (SES),
and
the quality of the SES triangulation is critical in boundary element simulations 
of the PB model.
In this work we compare the MSMS and NanoShaper surface triangulation codes 
for a set of 38 biomolecules.
While MSMS produces triangles of exceedingly small area
and
large aspect ratio,
the two codes yield comparable values for the SES surface area
and
electrostatic solvation energy,
where the latter calculations were performed using
the treecode-accelerated boundary integral (TABI) PB solver.
However we found that Nanoshaper is more efficient and reliable than MSMS,
especially when parameters are set to produce highly resolved triangulations. 
\end{abstract}

\begin{wileykeywords} solvated biomolecule, solvent excluded surface, electrostatics, Poisson--Boltzmann, boundary element method, treecode
\end{wileykeywords}

%

  \makeatletter
  \renewcommand\@biblabel[1]{#1.}
  \makeatother

\bibliographystyle{apsrev}

\renewcommand{\baselinestretch}{1.5}
\normalsize

\clearpage

\section*{\sffamily \Large INTRODUCTION}

\begin{figure}[htb]
    \centering
    \includegraphics[width=0.75\textwidth]{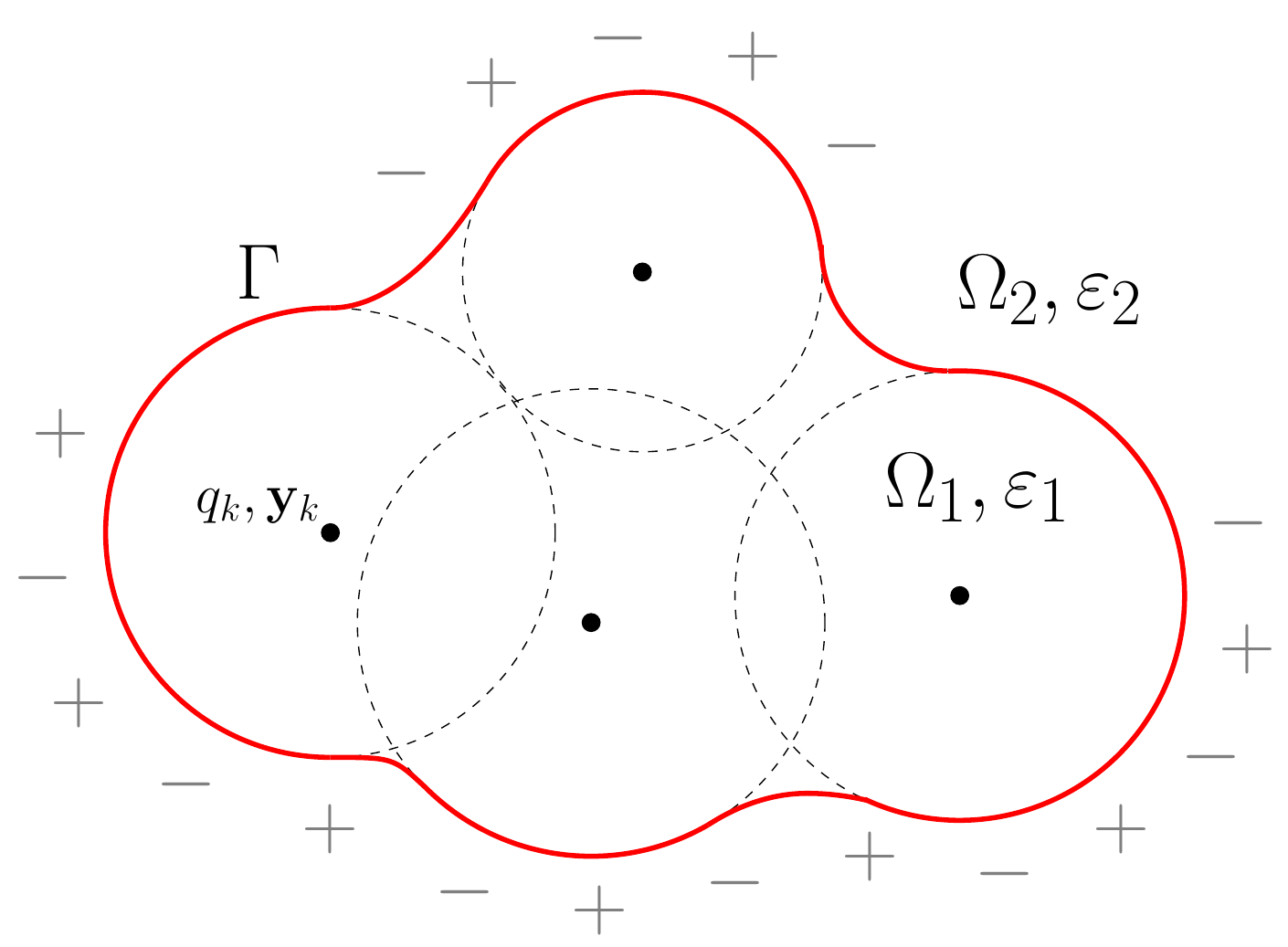}
    \caption{Poisson--Boltzmann implicit solvent model,
solute domain $\Omega_1$ with dielectric constant $\varepsilon_1$,
atomic charges $q_k$ located at ${\bf y_k}$,
vdW radii (dashed circles),
solvent domain $\Omega_2$ with dielectric constant $\varepsilon_2$,
dissolved salt ions $(+,-)$,
dielectric interface ($\Gamma$).}
    \label{PB_schematic}
\end{figure}

Implicit solvent models play a key role in computational modeling of electrostatic interactions between biomolecules and their solvent environment~\cite{Roux:1999aa, Zhang:2011aa, Tomasi:2004aa}. 
Of particular importance is the Poisson--Boltzmann (PB) implicit solvent model~\cite{Baker:2004aa,Lu:2008aa}. 
Figure~\ref{PB_schematic} shows
the interior domain $\Omega_1 \subset \mathbb{R}^3$ containing the solute biomolecule, 
the exterior domain $\Omega_2 = \mathbb{R}^3 \setminus \overline{\Omega}_1$ 
containing the ionic solvent,
and
the dielectric interface $\Gamma = \overline{\Omega}_1 \cap \overline{\Omega}_2$.
In a 1:1 electrolyte at low ionic concentration, 
one can utilize the linearized PB equation for the electrostatic potential $\phi$,
\begin{equation}
-\nabla\cdot\left(\varepsilon({\bf x})\nabla \phi({\bf x})\right) +  
\overline{\kappa}^2({\bf x})\phi({\bf x}) = 
\sum\limits_{k=1}^{N_c}q_k \delta\left({\bf x} - {\bf y}_k\right), \quad {\bf x} \in \mathbb{R}^3,
\label{eq:lpb}
\end{equation}
where $\varepsilon({\bf x})$ is the dielectric constant,
$\overline{\kappa}$ is the modified Debye-H\"{u}ckel inverse length in units of {\AA}$^{-1}$,
$N_c$ is the number of atoms in the solute biomolecule,
${\bf y}_k$ is the position of the $k$th solute atom,
and
$q_k$ is the associated partial charge in units of fundamental charge $e_c$.
The dielectric interface conditions are
\begin{equation}
\phi_1({\bf x}) = \phi_2({\bf x}), \quad
\varepsilon_1 \frac{\partial \phi_1({\bf x})}{\partial n} = 
\varepsilon_2 \frac{\partial \phi_2({\bf x})}{\partial n}, 
\quad{\bf x} \in \Gamma,
\label{eq:interface_conditions}
\end{equation}
where $\phi_1({\bf x})$ and $\phi_2({\bf x})$ are the limiting values 
approaching the interface $\Gamma$ 
from inside and outside the biomolecule, respectively, 
and $n$ indicates the outward normal direction on the interface. 
The first condition in Eq.~\eqref{eq:interface_conditions}
expresses continuity of the potential across the interface, 
and the second condition expresses continuity of the electric flux. 
The far-field boundary condition is
\begin{equation}
\lim_{|{\bf x}| \to \infty}\phi({\bf x}) = 0.
\end{equation}

The present work assumes that $\varepsilon$ and $\overline{\kappa}$ are piecewise constant,
\begin{equation}\label{eq:debyehuckel}
\varepsilon({\bf x}) =
\begin{cases}
\varepsilon_1, & {\bf x} \in \Omega_1, \\
\varepsilon_2, & {\bf x} \in \Omega_2,
\end{cases}
\quad , \quad
\overline{\kappa}^2({\bf x}) = 
\begin{cases} 
0, & {\bf x} \in \Omega_1, \\
\displaystyle
\left(\frac{8\pi N_Ae_c^2}{1000 k_BT}\right)I_s, & {\bf x} \in \Omega_2,
\end{cases}\end{equation}
where $N_A$ is Avogadro's number, 
$k_B$ is the Boltzmann constant, 
$T$ is the temperature,
and $I_s$ is the molar concentration of the ionic solvent. 
A key quantity of interest in analyzing protein stability
is the electrostatic solvation energy,
\begin{equation}
E_{\text{sol}} = 
\frac{1}{2}\sum\limits_{k=1}^{N_c}q_k\phi_{\rm reac}({\bf y}_k),
\label{eq:solveng}
\end{equation}
where the reaction field potential at an atomic position,
\begin{equation}
\phi_{\rm reac}({\bf y}_k) = 
\lim_{{\bf x} \to {\bf y}_k}\left(\phi({\bf x}) - \sum_{j=1}^{N_c}\frac{q_j}{4\pi |{\bf x}-{\bf y}_j|}\right),
\end{equation}
is the difference between the total potential and the Coulomb potential.

A variety of numerical methods have been applied to the PB model, 
including finite-difference \cite{Holst:1995aa, Baker:2001aa, Luo:2002aa, Wang:2010aa, Chen:2011ac, Boschitsch:2011aa, Geng:2012aa, Wilson:2016aa}, 
finite-element \cite{Lu:2008aa, Holst:2000aa, Baker:2000aa}, and 
boundary element~\cite{Liang:1997aa, Juffer:1991aa, Boschitsch:2002aa, Geng:2013aa, Cooper:2014aa} methods. 
The present work is concerned with boundary element methods (BEM)  
which solve for the surface potential on a triangulation of the interface;
these schemes benefit from rigorous enforcement of the interface conditions 
and the far-field boundary condition, 
but they face the difficulty of evaluating singular integrals
and
the expense of solving a dense linear system.
The treecode-accelerated boundary integral PB solver (TABI-PB~\cite{Geng:2013aa}) 
addresses these issues using a simple collocation scheme to discretize the integrals
and
a treecode algorithm to reduce the cost of solving the linear system
from $O(N^2)$ to $O(N \log N)$,
where $N$ is the number of triangles representing the interface.

\section*{\sffamily \Large Dieletric Interface}

Several models have been used for the dielectric interface 
between the solute and solvent 
in implicit solvent simulations~\cite{Chen:2011aa,Decherchi:2013aa}. 
The {\bf van der Waals (vdW) surface}, the simplest of these models, 
is the union of hard spheres with vdW radii
representing the atoms comprising the biomolecule. 
The {\bf solvent accessible surface} (SAS) is formed by tracing the center of a 
probe sphere representing a water molecule rolling along the exterior of the vdW surface;
the SAS surface is equivalent to a vdW surface in which the vdW radii are increased by the
probe sphere radius. 
The {\bf solvent excluded surface} (SES) is formed by the inward facing surface 
of the probe sphere rolling along the vdW surface~\cite{Richards:1977aa, Connolly:1985aa}. 
The SES surface is comprised of spherical contact patches 
where the probe sphere touches the vdW surface, 
and 
toroidal reentrant patches formed by the inward facing surface of the probe sphere
when it does not touch the vdW surface, i.e., 
when it is in contact with more than one solute atom.
The {\bf skin surface}~\cite{Rocchia:2002aa,Rocchia:2005aa,Cheng:2009aa}
is comprised of spherical and hyperboloid patches
constructed from a set of spheres through shrinking and convex combinations. 
The {\bf Gaussian surface}~\cite{Chen:2012aa} is the level set 
of a linear combination of Gaussian functions centered at the solute atoms.

Several algorithms have been developed for triangulating these surfaces, 
where the input is the location and radii of the solute atoms
and
the output is a list of triangles that triangulate the surface.
An alternative approach uses a level-set representation of the surface
in an adaptive Cartesian grid~\cite{Can:2006aa,Egan:2018aa}.
Publicly available surface triangulation codes include
MSMS~\cite{Sanner:1995aa},
EDTSurf~\cite{Xu:2009aa, Xu:2013aa},
TMSmesh~\cite{Chen:2011aa, Chen:2012aa}, 
and
NanoShaper~\cite{Decherchi:2013ab}.
Previous work investigated the performance of 
SES surfaces and skin surfaces in the finite-difference Delphi code~\cite{Decherchi:2013ab},
and 
the performance of Gaussian surfaces relative to SES surfaces in 
the boundary element fast multipole code AFMPB~\cite{Liu:2015aa}.
The present work focuses on the SES surface 
and
compares the performance of
MSMS and NanoShaper in computing the surface area
and electrostatic solvation energy within the boundary element TABI-PB framework.

\subsection*{\sffamily \large MSMS} 
MSMS, introduced by Sanner in 1995~\cite{Sanner:1995aa}, 
has gained widespread popularity for generating SES surface triangulations.
After creating an analytical representation of the surface, 
the algorithm generates a triangulation of specified density by fitting predefined triangulated patches to the surface. 
The mesh resolution is controlled by the user-specified density parameter $d$ 
that sets the number of triangles in the triangulation of a given surface 
in units of vertices/angstrom$^2$.

\subsection*{\sffamily \large NanoShaper}
NanoShaper, introduced by Decherchi and Rocchia in 2012~\cite{Decherchi:2013aa}, 
implements the SES surface as well as several alternatives including the
Gaussian and skin surfaces. 
In constructing an SES surface triangulation, 
NanoShaper first builds a description of the surface with a set of patches, 
analytically if possible or else with an approximation.
The code then employs a ray-casting algorithm in which rays parallel to the 
coordinate axes are cast and intersections with the surface are calculated. 
The vertex positions of intersection are then used by the marching cubes algorithm to obtain the triangulation. 
The mesh resolution is controlled by the user-specified scaling parameter $s$ 
that sets the inverse side length of a cubic grid cell in units of angstroms.

The SES surface triangulations are commonly used to compute molecular properties
such as the surface area,
electrostatic potential,
and
electrostatic solvation energy of biomolecules.
The present work investigates the triangulations produced by 
MSMS and NanoShaper 
in terms of their effect on the accuracy and efficiency of these computations.

\section*{\sffamily \Large The TABI-PB Solver}

The TABI-PB solver~\cite{Geng:2013aa}
relies on a reformulation of the PB Eq.~\eqref{eq:lpb}
developed by Juffer et~al. as a coupled set of boundary integral equations
for the surface potential $\phi_1$ 
and its normal derivative $\partial\phi_1/\partial n$ 
on the dielectric interface~\cite{Juffer:1991aa},
\begin{subequations}\label{eq:integraleq}
\begin{align}
\frac{1}{2}\left(1+\varepsilon\right)\phi_1\left({\bf x}\right) 
=& \int_\Gamma\left[K_1\left({\bf x}, {\bf y}\right) \frac{\partial \phi_1\left({\bf y}\right)}{\partial n} + K_2\left({\bf x}, {\bf y}\right) \phi_1\left({\bf y}\right) \right]dS_{\bf y} + S_1({\bf x}), 
\quad {\bf x} \in \Gamma, \\
\frac{1}{2}\left(1+\frac{1}{\varepsilon}\right) \frac{\partial \phi_1\left({\bf x}\right)}{\partial n}
=& \int_\Gamma\left[K_3\left({\bf x}, {\bf y}\right) \frac{\partial \phi_1\left({\bf y}\right)}{\partial n} + K_4\left({\bf x}, {\bf y}\right) \phi_1\left({\bf y}\right) \right]dS_{\bf y} + S_2({\bf x}),
\quad {\bf x} \in \Gamma,
\end{align}
\end{subequations}
where $\varepsilon = \varepsilon_1 / \varepsilon_2$ 
is the solute/solvent ratio of dielectric constants. 
The kernels $K_1, K_2, K_3, K_4$ depend on the Coulomb and screened Coulomb potentials,
\begin{equation}
G_0\left({\bf x}, {\bf y}\right) = \frac{1}{4\pi \left|{\bf x} - {\bf y}\right|}, \ \ 
G_\kappa\left({\bf x}, {\bf y}\right) = \frac{e^{-\kappa\left|{\bf x} - {\bf y}\right|}}{4\pi \left|{\bf x} - {\bf y}\right|}.
\end{equation}
and
the source terms are
\begin{equation}
S_1({\bf x}) = \frac{1}{\varepsilon_1}\sum\limits_{k=1}^{N_c}q_k G_0\left({\bf x}, {\bf y}_k\right), \ \ 
S_2({\bf x}) = \frac{1}{\varepsilon_1}\sum\limits_{k=1}^{N_c}q_k \frac{\partial G_0\left({\bf x}, {\bf y}_k\right)}{\partial n_{\bf x}}.
\end{equation}

In this context the electrostatic solvation energy in Eq.~\eqref{eq:solveng} is
obtained from the surface potential and its normal derivative, 
\begin{equation}
E_{\text{sol}} = 
\frac{1}{2}\sum\limits_{k=1}^{N_c}q_k\int_\Gamma\left[K_1({\bf y}_k,{\bf y})  
\frac{\partial \phi_1\left({\bf y}\right)}{\partial n} + 
K_2({\bf y}_k,{\bf y})\phi_1({\bf y})\right]dS_{\bf y}.
\label{eq:solvengboundary}
\end{equation}

The TABI-PB solver calculates the surface integrals using a boundary element method 
on the triangulated SES surface,
where the collocation points are the triangle centroids.
The resulting linear system for the surface potentials
and
their normal derivatives is solved by GMRES iteration,
while a treecode algorithm is employed to accelerate the matrix-vector product
in each step of the iteration~\cite{Geng:2013aa}.

\section*{\sffamily \Large METHODOLOGY}

To assess the SES surface triangulations produced by MSMS and NanoShaper,
we compute the surface area $S_\text{a}$ and electrostatic solvation energy $E_\text{sol}$ 
for a test set of 38~biomolecules comprising peptides, proteins, and nucleic acid fragments,
where 
$S_{\rm a}$ is computed by summing the areas of the triangles
and
$E_\text{sol}$ is computed using TABI-PB. 
In addition to examining the accuracy of these results,
we report the total CPU time for the TABI-PB computation;
the CPU time for generating the triangulation and other pre-processing steps
is negligible in comparison to the boundary element computation time.
Surface visualizations were generated with VTK ParaView.

The biomolecules in the test set, listed in Table \ref{table.testset}, 
are those with widely available PDB entries from the list used in a previous 
molecular surface comparison study\cite{Liu:2015aa}. 
We generate PQR files for each test biomolecule using PDB2PQR \cite{Dolinsky:2004aa} 
with the CHARMM force field and water molecules removed.

\begin{table}[htb]
\caption{PDB ID and number of atoms for test set of 38 biomolecules
comprising proteins, peptides, and nucleic acid fragments~\cite{Liu:2015aa}.}
\label{table.testset}
\begin{center}
\begin{tabular}{lllllllll}
\hline
Index     	& $1$ 	& $2$	& $3$ 	& $4$ 	& $5$ 	& $6$ 	& $7$ 	& $8$ \\
PDB ID     & 2LWC	& 1GNA	& 1S4J	& 1CB3	& 1V4Z 	& 1BTQ 	& 1I2X 	& 1AIE \\
\# atoms   	& $75$ 	& $163$ 	& $182$ 	& $183$	& $266$ 	& $304$ 	& $513$ 	& $522$ \\
\hline
Index	& $9$	& $10$ 	& $11$ 	& $12$	& $13$ 	& $14$	& $15$ 	& $16$       \\
PDB ID	& 1ZWF	& 375D	& 440D 	& 4HLI 	& 3ES0	& 3IM3 	& 2IJI 	& 1COA 	 \\
\# atoms	& $586$	& $593$	& $629$ & $697$ 	& $781$	& $851$ 	& $890$	& $1057$ 	   \\
\hline
Index	& $17$	& $18$ 	& $19$ 	& $20$	& $21$ 	& $22$	& $23$ 	& $24$       \\
PDB ID	& 2AVP	& 1SM5	& 2ONT 	& 4GSG	& 3ICB	& 1DCW	& 3LDE 	& 1AYI 	  \\
\# atoms	& $1085$	& $1137$	& $1161$	& $1195$	& $1202$	& $1257$	& $1294$ 	& $1365$	   \\
\hline
Index	& $25$	& $26$ 	& $27$ 	& $28$	& $29$ 	& $30$	& $31$ 	& $32$       \\
PDB ID	& 2YX5	& 3DFG	& 3LOD 	& 1TR4	& 1RMP	& 1IF4	& 4DUT 	& 3SQE 	 \\
\# atoms	& $1385$	& $2198$	& $2246$ & $3423$ 	& $3478$	& $4071$	& $4217$ 	& $4647$	  \\
\hline
Index	& $33$	& $34$ 	& $35$ 	& $36$	& $37$ 	& $38$	& 	&       \\
PDB ID	& 1HG8	& 4DPF	& 3FR0 	& 2H8H 	& 2CEK	& 1IL5 	& 	&  	  \\
\# atoms	& $4960$ & $5824$	& $6952$ & $7084$ 	& $8346$	& $8349$ 	& 	&  	    \\
\hline
\end{tabular}
\end{center}
\end{table}

The MSMS triangulations were generated using density values $d = 1, 2, 4, 8, 16$
and 
the NanoShaper triangulatons were generated using scaling parameter values $s = 1, 2, 3, 4, 5$.
For all surfaces a probe radius of $1.4$~\AA~was used.

The physical parameter values were 
ionic concentration $I_s = 0.15$ M, 
temperature $T = 300$ K, 
and
solute and solvent dielectric constants $\varepsilon_1 = 1$, $\varepsilon_2 = 80$.
The treecode parameters were
multipole acceptance criterion $\theta = 0.8$,
Taylor series order $p=3$,
and
maximum number of particles in a leaf $N_0 = 500$.
The GMRES tolerance was 1E-4,
with 10 iterations between restarts
and
maximum number of iterations 110. 

All computations were performed in serial on the University of Michigan FLUX cluster, 
with Intel Xeon CPUs running at either 2.5 or 2.8\,GHz. 
In this system the exact processors could not be specified, 
so the timing results were averaged over multiple runs. 
The code was compiled with gfortran using the -O2 optimization flag.
The newest version of the TABI-PB solver is available on GitHub at \texttt{github.com/Treecodes/TABI-PB}. 
The version of the TABI-PB solver used in this work is available on GitHub at \texttt{github.com/lwwilson/TABI-PB}.
We also note here that TABI-PB was recently implemented 
as an option in the APBS package developed at Pacific Northwest National Laboratory~\cite{Jurrus:2017aa}.

\section*{\sffamily \Large RESULTS AND DISCUSSION}

We first study geometric features of the surface triangulations 
by considering the 
triangle size, shape, and aspect ratio,
and 
qualitatively comparing the generated surface meshes.
We then extrapolate with respect to the number of triangles 
to calculate highly accurate converged values of the surface area and solvation energy; 
the converged result is the $y$-intercept of a simple extrapolation of  
the computed surface area or solvation energy versus $N^{-1}$ for the 
two highest resolution meshes produced by MSMS and NanoShaper. 
For some values of the density parameter $d$,
MSMS produced spurious results for the larger molecules or 
failed to even produce a triangulation at all,
as observed by previous investigators~\cite{Can:2006aa}; 
in these cases the extrapolation used the highest resolution meshes 
for which MSMS did not fail.

\section*{\sffamily \Large Triangulation Filter}

Both MSMS and NanoShaper produce a number of small or thin triangles
which reduce the computational accuracy and efficiency, 
and 
it is common practice to delete them from the simulations.
Hence in the present work,
triangles are deleted if their area is less than 1e--5\,\AA$^2$
or 
if the distance between the centroids of two neighboring triangles is less than 1e--5\,\AA.
Table~\ref{table:avgnontriangles} gives the percent of deleted triangles averaged over all
triangulations using MSMS and NanoShaper.
Among the deleted triangles,
some had area less than machine precision and these are designated as zero-area triangles.
With MSMS the deleted triangles are 0.064\,\% of the total,
while with NanoShaper the total is more than 100 times smaller.
Table~\ref{table:avgnontriangles} 
also shows that most of the deleted MSMS triangles were zero-area,
while none of this type were produced by NanoShaper.

\begin{table}[htb]
\caption{
Triangulation filter results showing percent of
deleted triangles and zero-area triangles,
values shown are averaged over all triangulations using MSMS and NanoShaper,
zero-area triangles are a subset of deleted triangles.} 
\label{table:avgnontriangles}
\begin{center}
\begin{tabular}{ccc}
\hline
code & deleted triangles & zero-area triangles \\
\hline
MSMS 		& 6.4e--2\,\% & 5.4e--2\,\% \\
NanoShaper	& 5.2e--4\,\% & 0.0e--0\,\% \\
\hline
\end{tabular}
\end{center} 
\end{table}

\section*{\sffamily \Large Triangle Aspect Ratios}

\begin{figure}[htb]
\centering
\begin{tabular}{cc}
\includegraphics[width=0.5\linewidth]{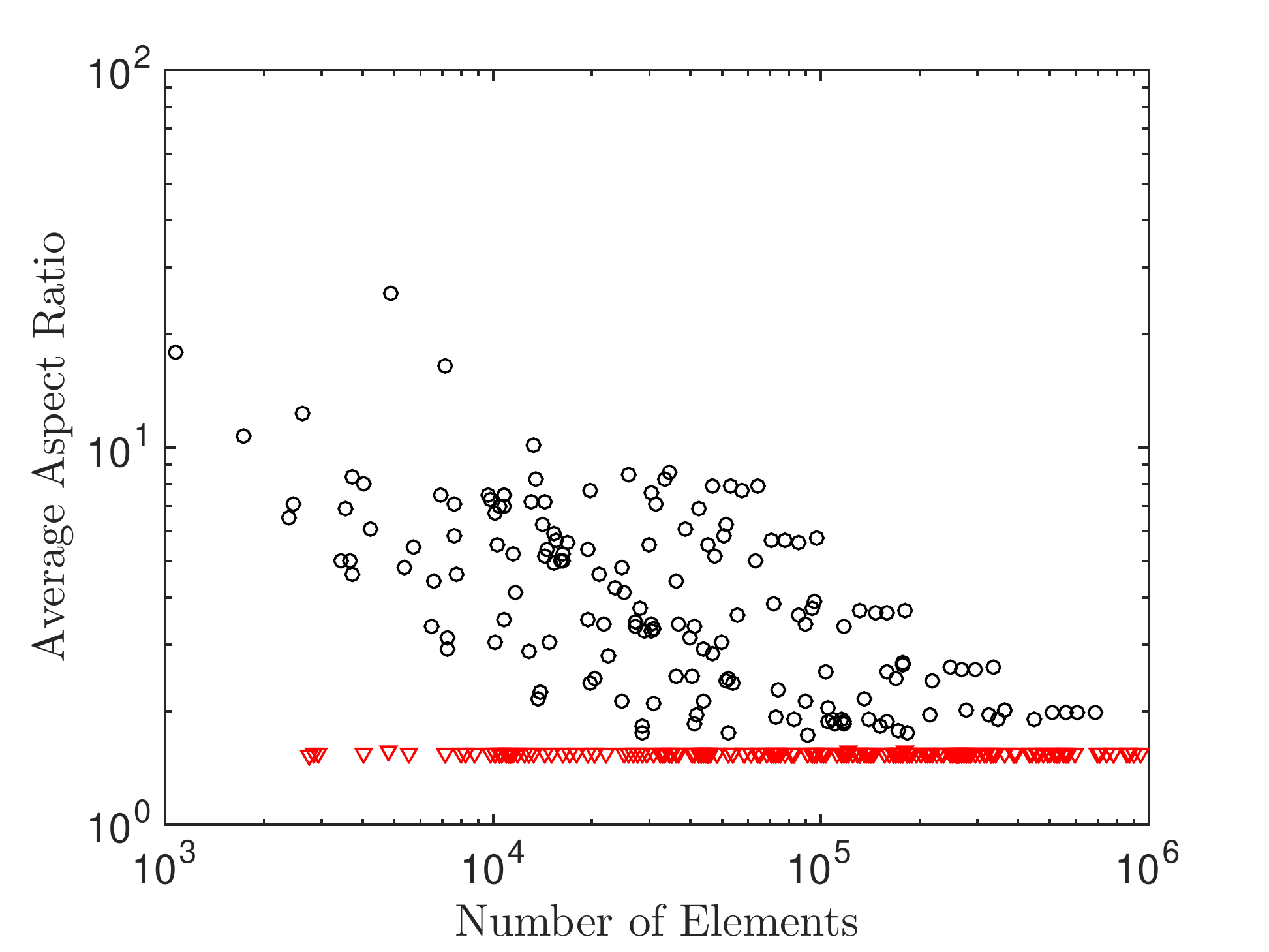} &
\includegraphics[width=0.5\linewidth]{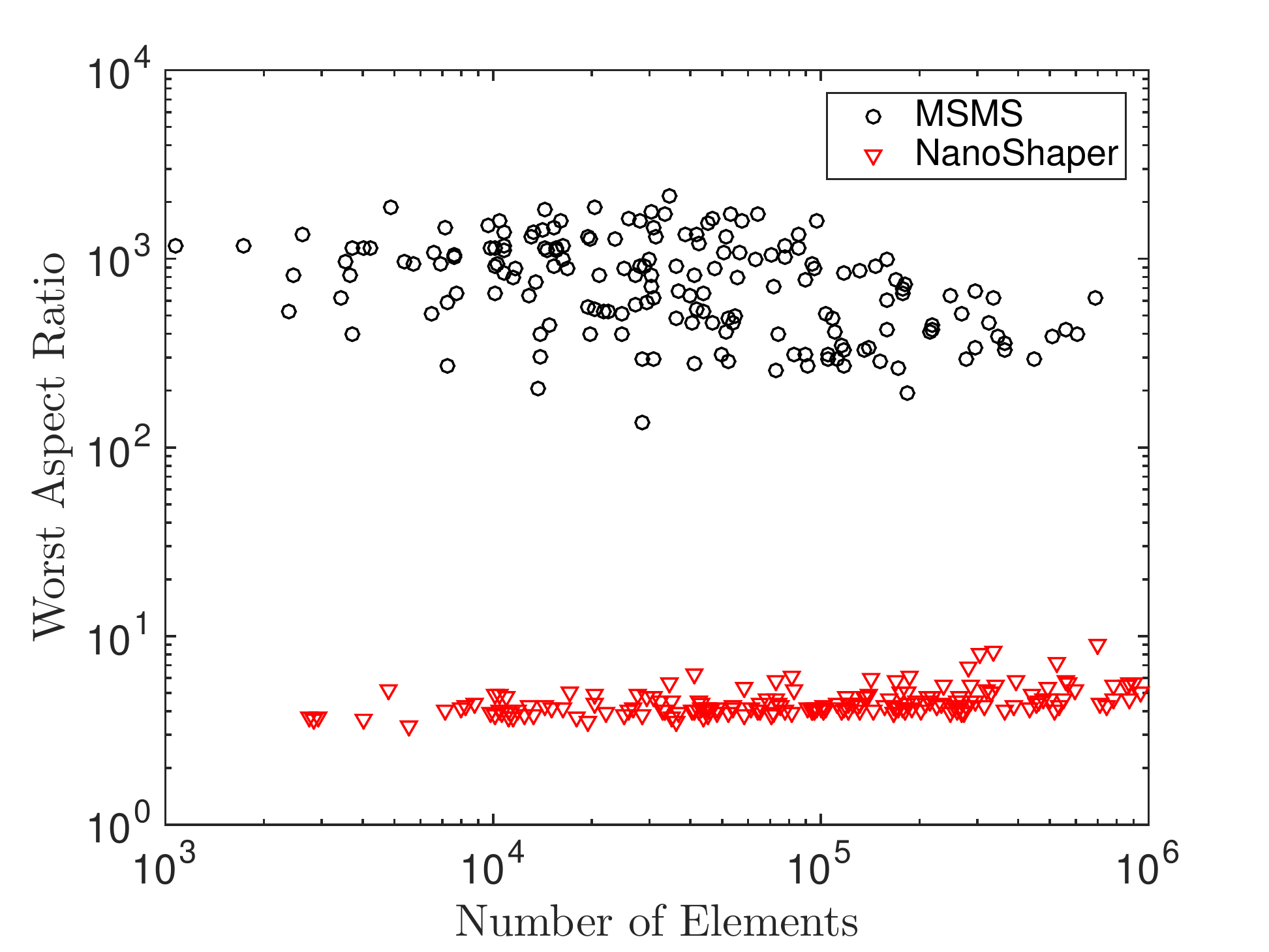} \\
(a) average aspect ratio, $r_{avg}$ & (b) maximum aspect ratio, $r_{max}$ \\
\end{tabular}
\caption{
Triangle aspect ratio versus number of elements $N$ for each generated surface,
(a) average aspect ratio, $r_{avg}$,
(b) maximum aspect ratio, $r_{max}$,
MSMS ($\circ$, black), NanoShaper ($\triangledown$, red).}
\label{fig:aspectratio}
\end{figure}

Even after filtering the triangulation as above,
the aspect ratio of the remaining triangles can affect the 
computational performance.
The aspect ratio of a triangular surface element is defined as
the ratio of the longest to shortest sides.
Figure~\ref{fig:aspectratio} displays the (a) average aspect ratio, $r_{avg}$,
and 
(b) maximum aspect ratio, $r_{max}$, versus the number of triangles, $N$,
for each triangulation,
where $N$ varies from approximately 1e+3 to 1e+6,
for the chosen density and scaling parameters.
Figure~\ref{fig:aspectratio}a 
shows that the average aspect ratio of MSMS triangles is as large as $r_{avg} \approx 30$ for small $N$
and decreases to approximately $r_{avg} \approx 2$ for large $N$,
while the average aspect ratio of NanoShaper triangles is closer to $r_{avg} \approx 1$ for all $N$.
Figure~\ref{fig:aspectratio}b shows that the maximum aspect ratio of MSMS triangles 
varies between approximately 1e+2 and 2e+3,
while the maximum aspect ratio of NanoShaper triangles is below 1e+1
across all triangulations.

\section*{\sffamily \Large Surface Mesh Features}

\begin{figure}[htb]
\centering
\begin{tabular}{cccc}
\includegraphics[width=0.4\linewidth]{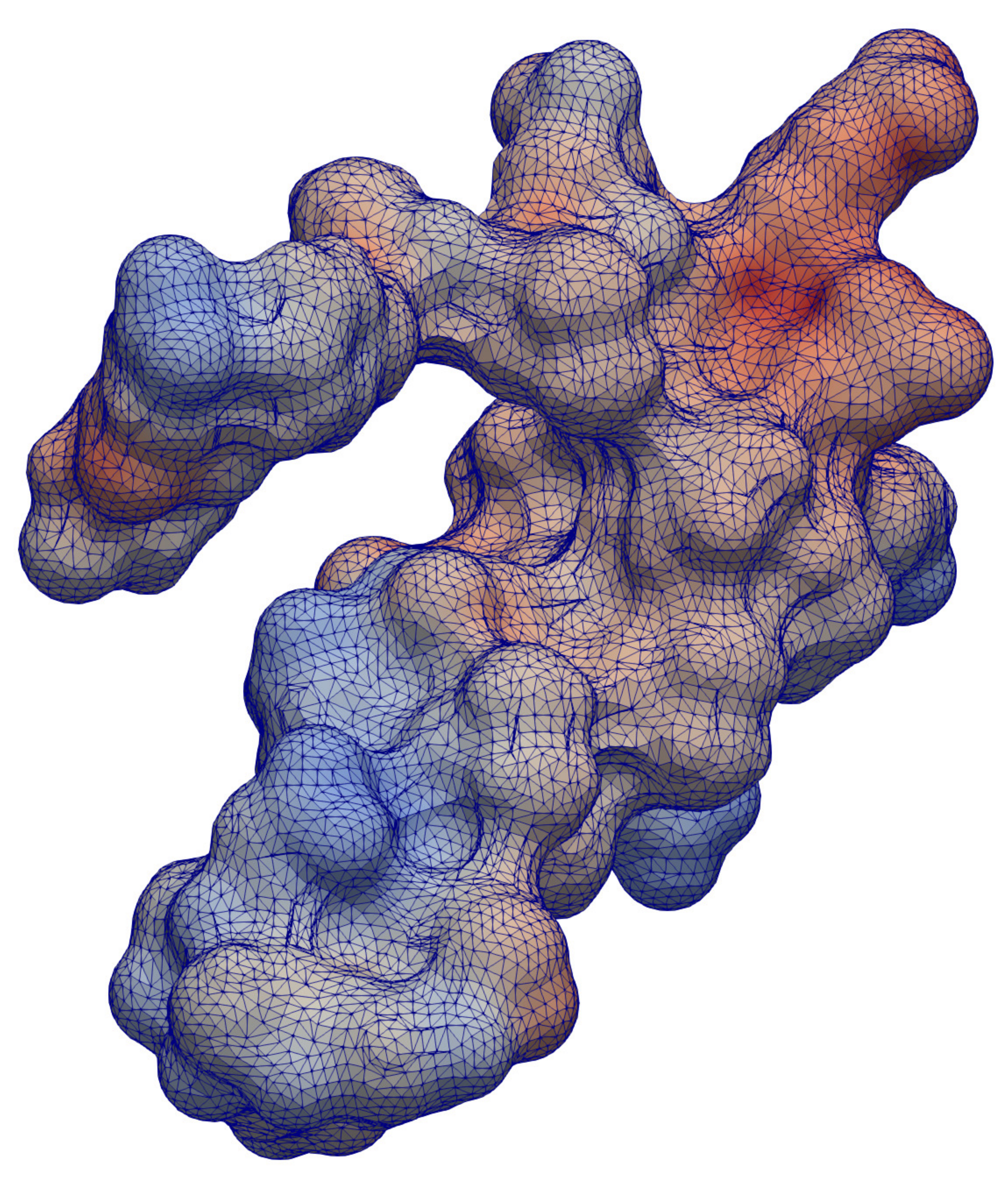} &
\includegraphics[width=0.4\linewidth]{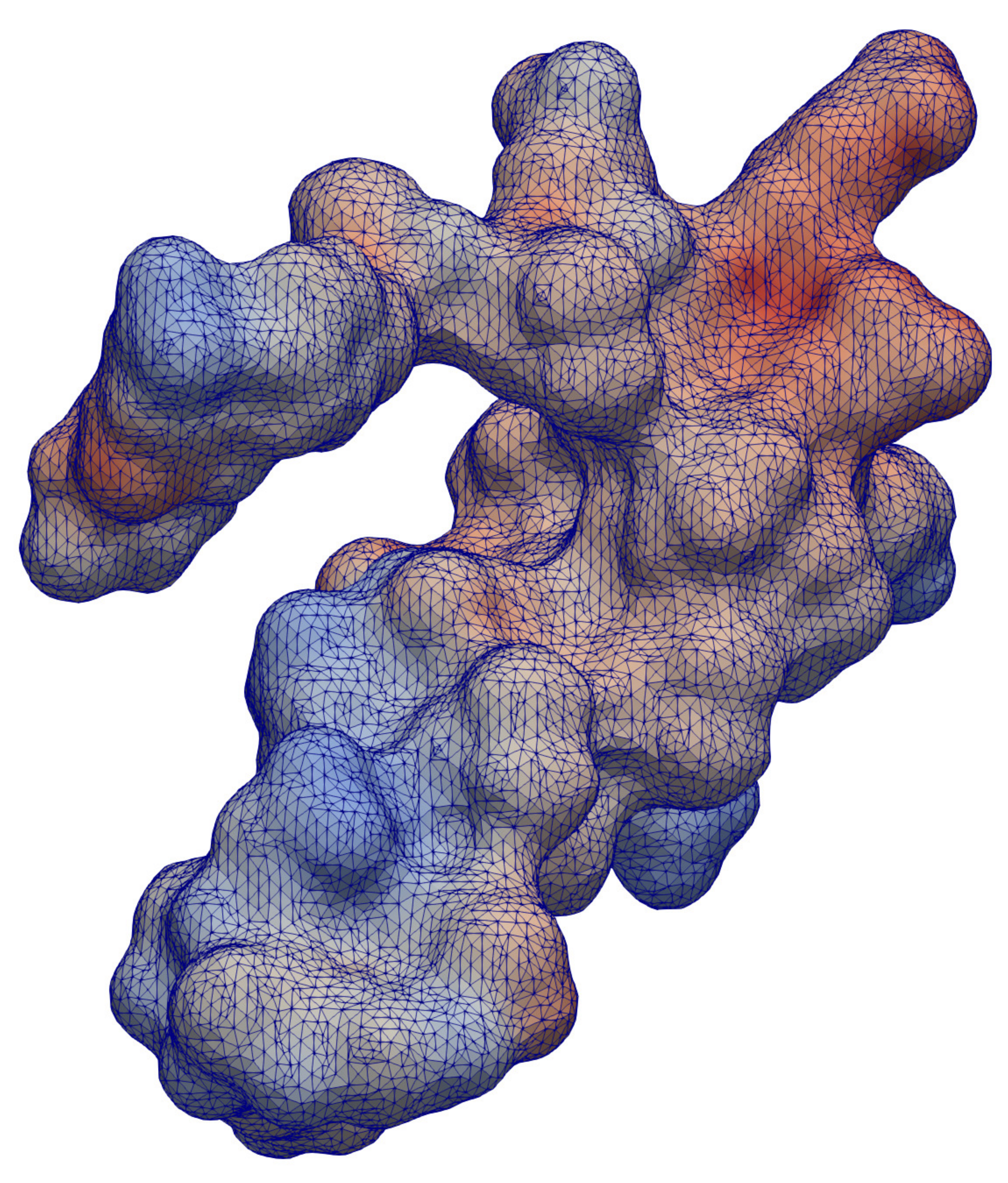} &
\includegraphics[width=0.15\linewidth]{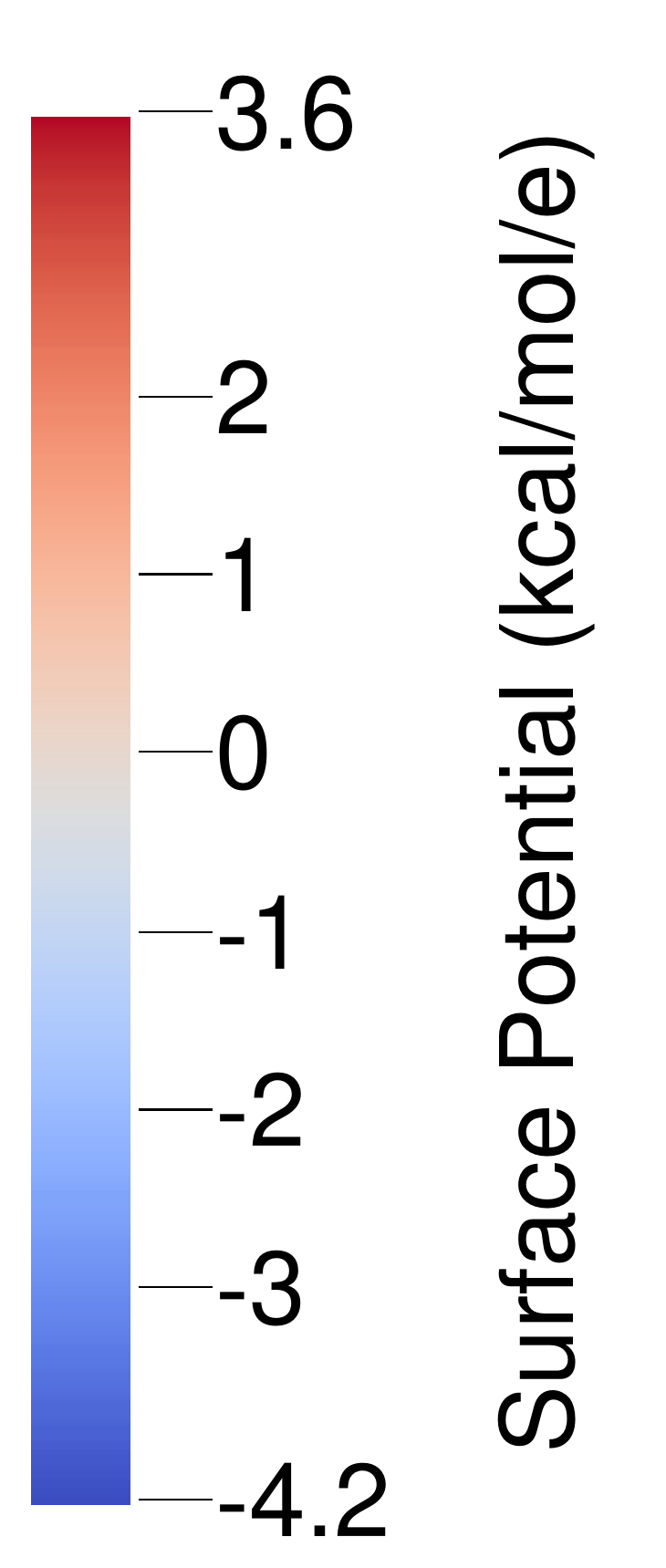} \\
(a) MSMS & (b) NanoShaper & \\ \\
\end{tabular}
\caption{
Protein 1AIE,
SES triangulation and electrostatic potential,
(a) MSMS, density $d=6$, $N=31480$ triangles, 
(b) NanoShaper, scaling parameter $s=2$, $N=32208$ triangles.}
\label{fig:surfacecomp1}
\end{figure}

\begin{figure}[htb]
\centering
\begin{tabular}{cccc}
  \begin{tikzpicture}
    \node[anchor=south west,inner sep=0] (image) at (0,0) {\includegraphics[width=0.4\textwidth]{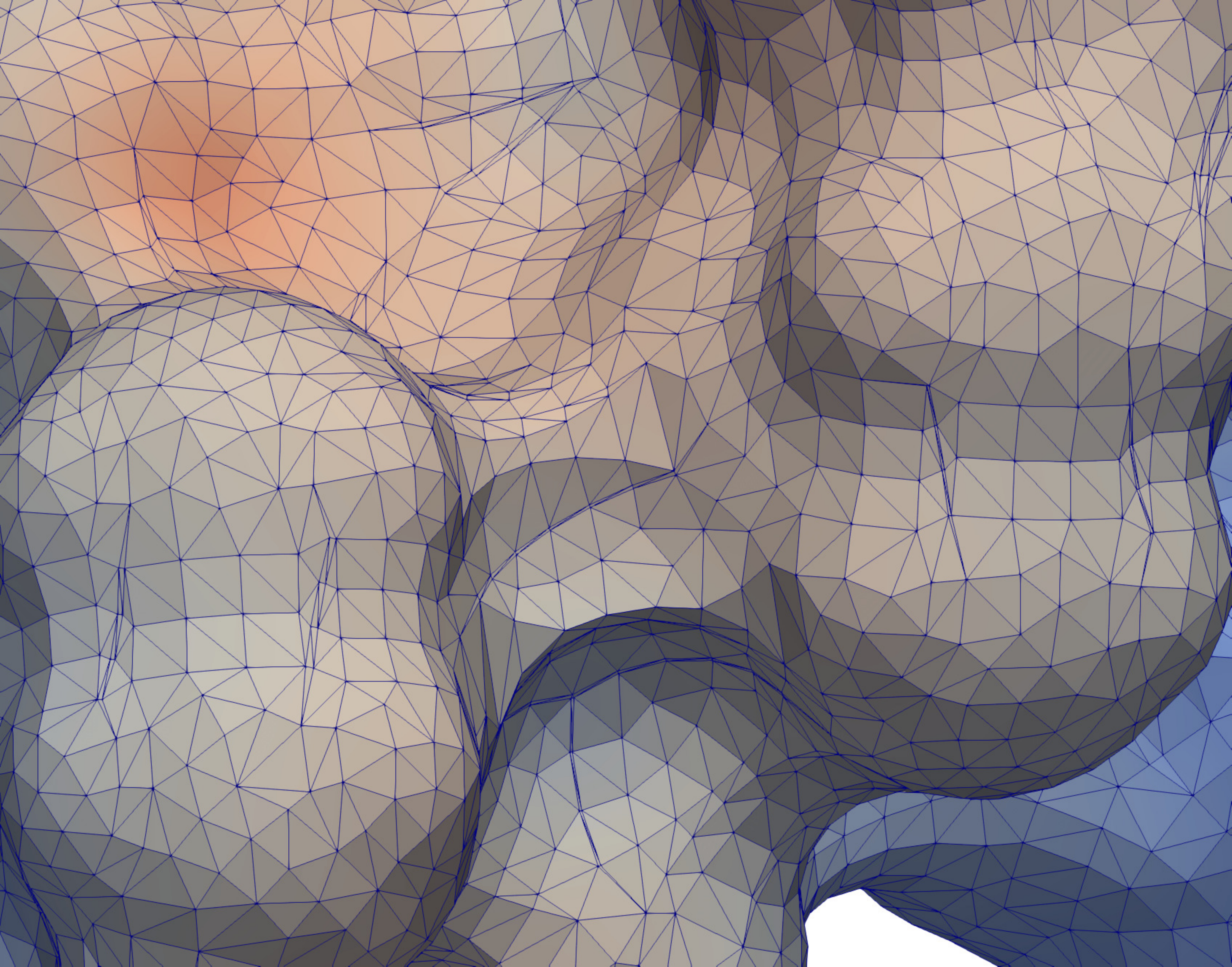}};
    \begin{scope}[x={(image.south east)},y={(image.north west)}]
        \draw[green,ultra thick,rounded corners] (0.07,0.25) rectangle (0.13,0.45);
        \draw[green,ultra thick,rounded corners] (0.72,0.43) rectangle (0.79,0.63);
        \draw[green,ultra thick,rounded corners] (0.42,0.08) rectangle (0.5,0.3);
        \draw[white,ultra thick,rounded corners] (0.42,0.33) rectangle (0.6,0.4);
        \draw[green,ultra thick,rounded corners] (0.23,0.26) rectangle (0.29,0.46);
        \draw[green,ultra thick,rounded corners] (0.88,0.43) rectangle (0.95,0.63);
        \draw[green,ultra thick,rounded corners] (0.3,0.84) rectangle (0.5,0.94);
    \end{scope}
  \end{tikzpicture} &
  \begin{tikzpicture} 
    \node[anchor=south west,inner sep=0] (image) at (0,0) {\includegraphics[width=0.4\textwidth]{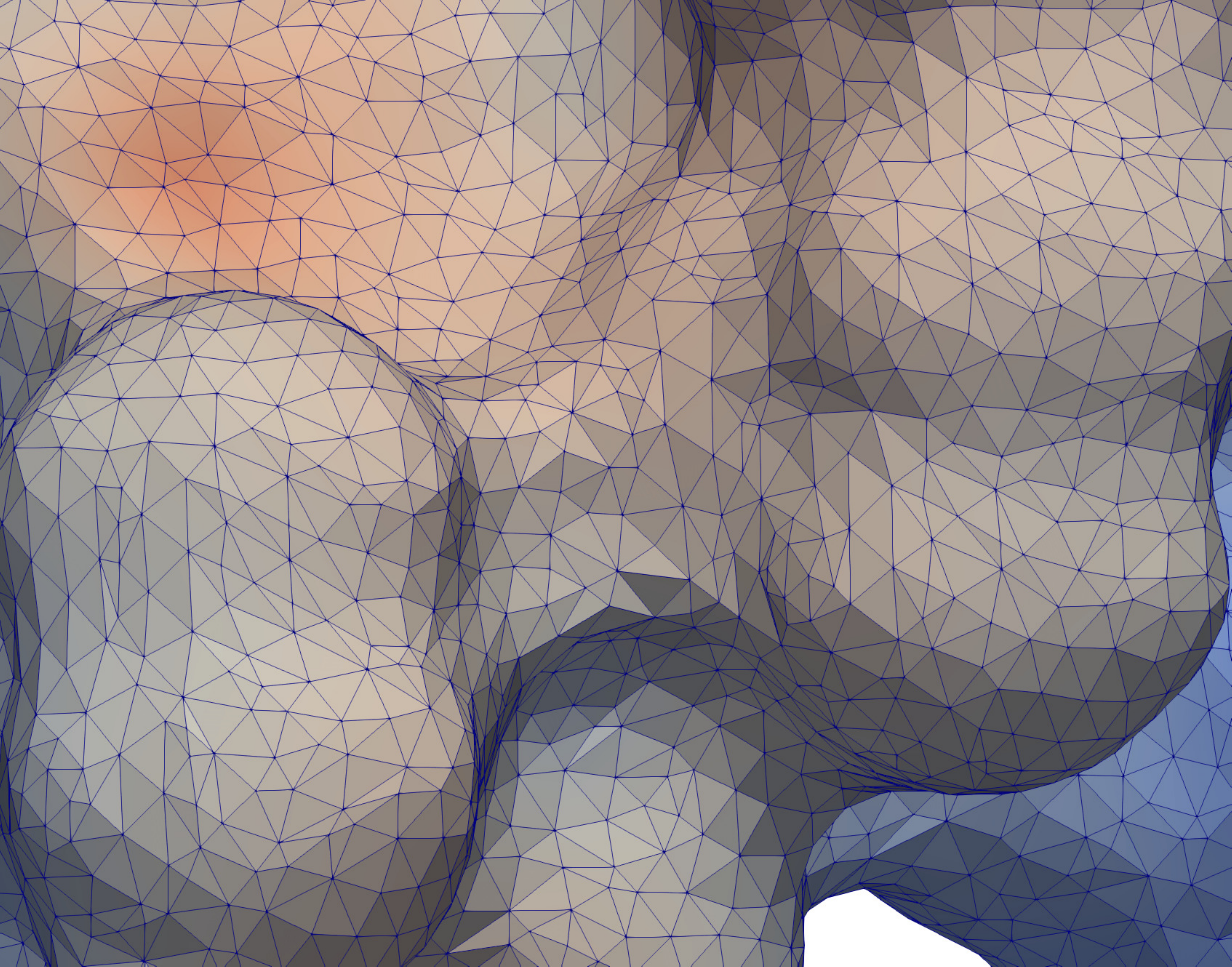}};
   \begin{scope}[x={(image.south east)},y={(image.north west)}]
        \draw[yellow,ultra thick,rounded corners] (0.45,0.33) rectangle (0.6,0.45);
    \end{scope} 
  \end{tikzpicture} & \\
  (a) MSMS, zoomed & (b) NanoShaper, zoomed & \\
\end{tabular}
\caption{
Protein 1AIE,
zoom of SES triangulation,
(a) MSMS, density $d=6$, $N=31480$ triangles, 
green boxes enclose {\it stitches} formed by high aspect ratio triangles, 
white box encloses a {\it cusp} formed by neighboring triangles 
that meet at an acute angle,
(b) NanoShaper, scaling parameter $s=2$, $N=32208$ triangles,
yellow box encloses a possible irregular feature.}
\label{fig:surfacecomp2}
\end{figure}

Figure~\ref{fig:surfacecomp1} displays the triangulation and surface potential
for a representative protein (1AIE) using (a) MSMS and (b) NanoShaper
with similar resolution,
$N \approx$ 3e+4 triangles in each case.
The surfaces are similar at first glance,
although the NanoShaper surface appears slightly smoother than the MSMS surface. 
Figure~\ref{fig:surfacecomp2} displays a zoom of the triangulations,
where several irregular features are highlighted;
in the MSMS mesh,
green boxes enclose {\it stitches} formed by high aspect ratio triangles,
and
a white box encloses a {\it cusp} formed by neighboring triangles that
meet at a acute angle,
while in the NanoShaper mesh,
a yellow box encloses a possible irregular feature,
which could in fact simply be an artifact of the surface lighting.
It should be noted that
irregular features are present in the MSMS mesh even after filtering;
by contrast, 
the NanoShaper mesh is relatively free of such irregular features.
The irregular features diminish the efficiency of the calculations;
as shown below, 
calculations using MSMS meshes require more iterations to converge
in comparison with 
calculations using NanoShaper meshes.

\section*{\sffamily \Large Dependence of $S_a$ and $E_{\rm sol}$ on Mesh Resolution}

\begin{figure}[htb]
\centering
\begin{tabular}{cc}
  \begin{tikzpicture}
    \draw (0, 0) node[inner sep=0] {\includegraphics[width=0.5\textwidth]{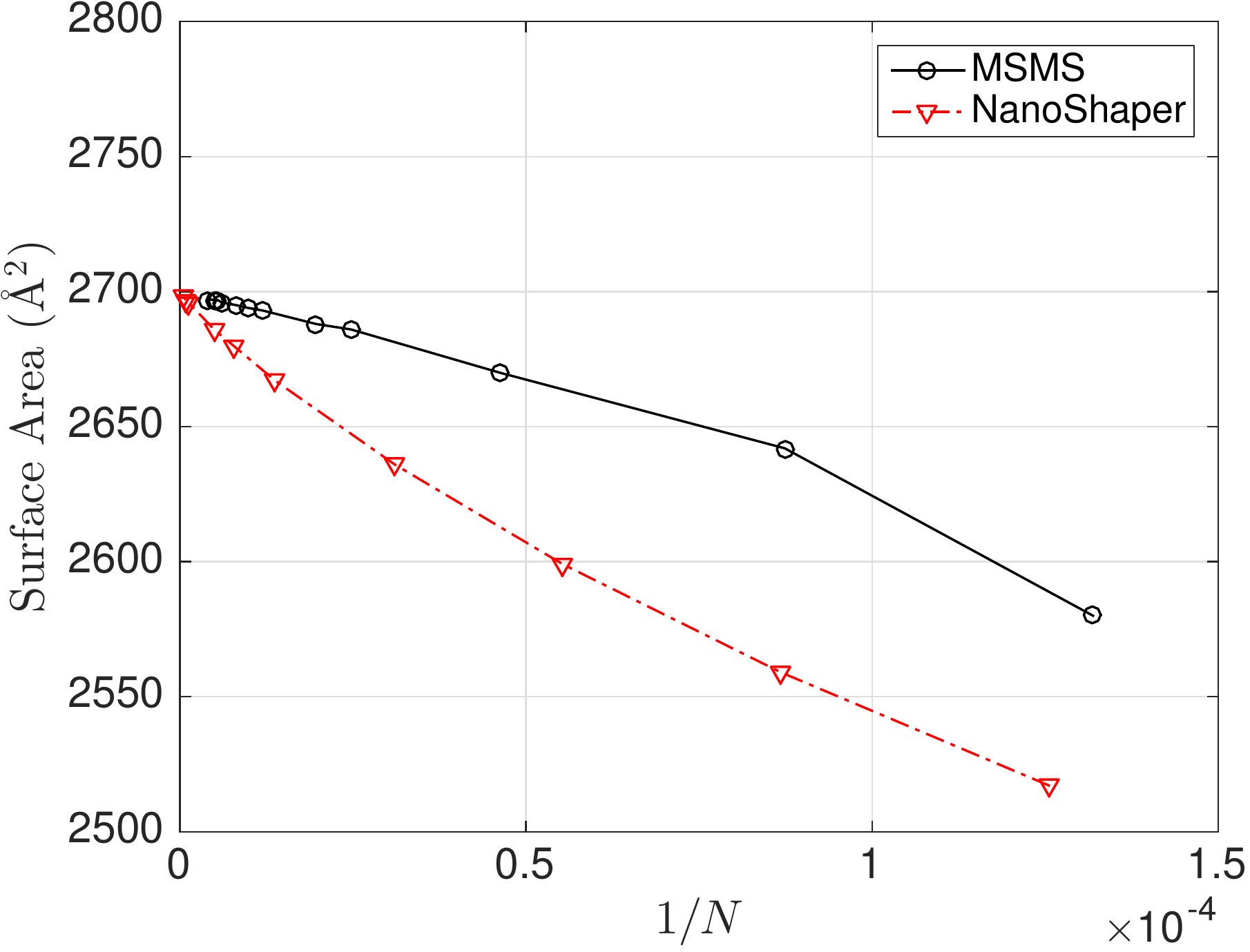}};
    \draw (-2, 2.5) node {(a) 1AIE};
  \end{tikzpicture} &
  \begin{tikzpicture}
    \draw (0, 0) node[inner sep=0] {\includegraphics[width=0.5\textwidth]{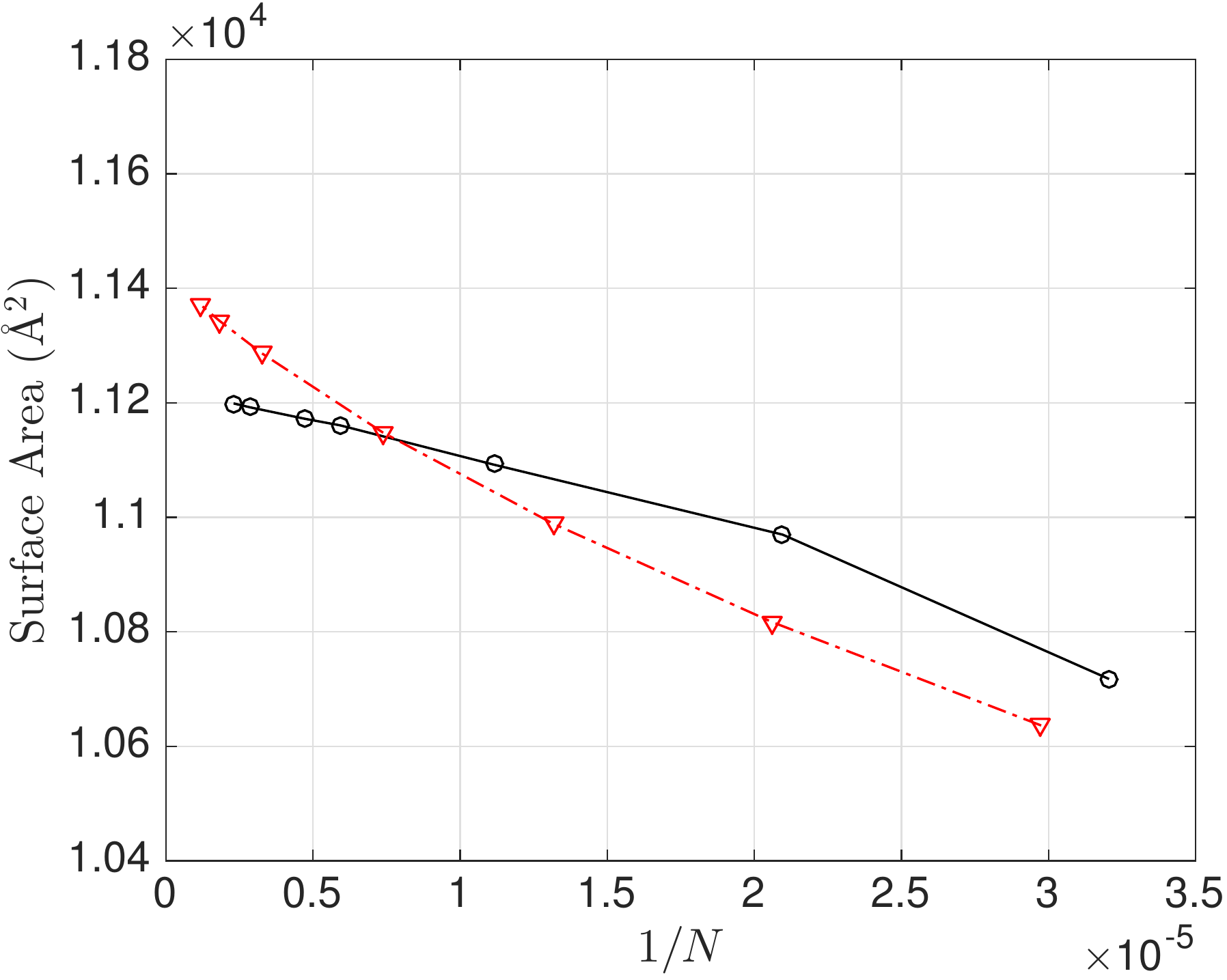}};
    \draw (-2, 2.5) node {(b) 1HG8};
  \end{tikzpicture} \\
  \begin{tikzpicture}
    \draw (0, 0) node[inner sep=0] {\includegraphics[width=0.5\textwidth]{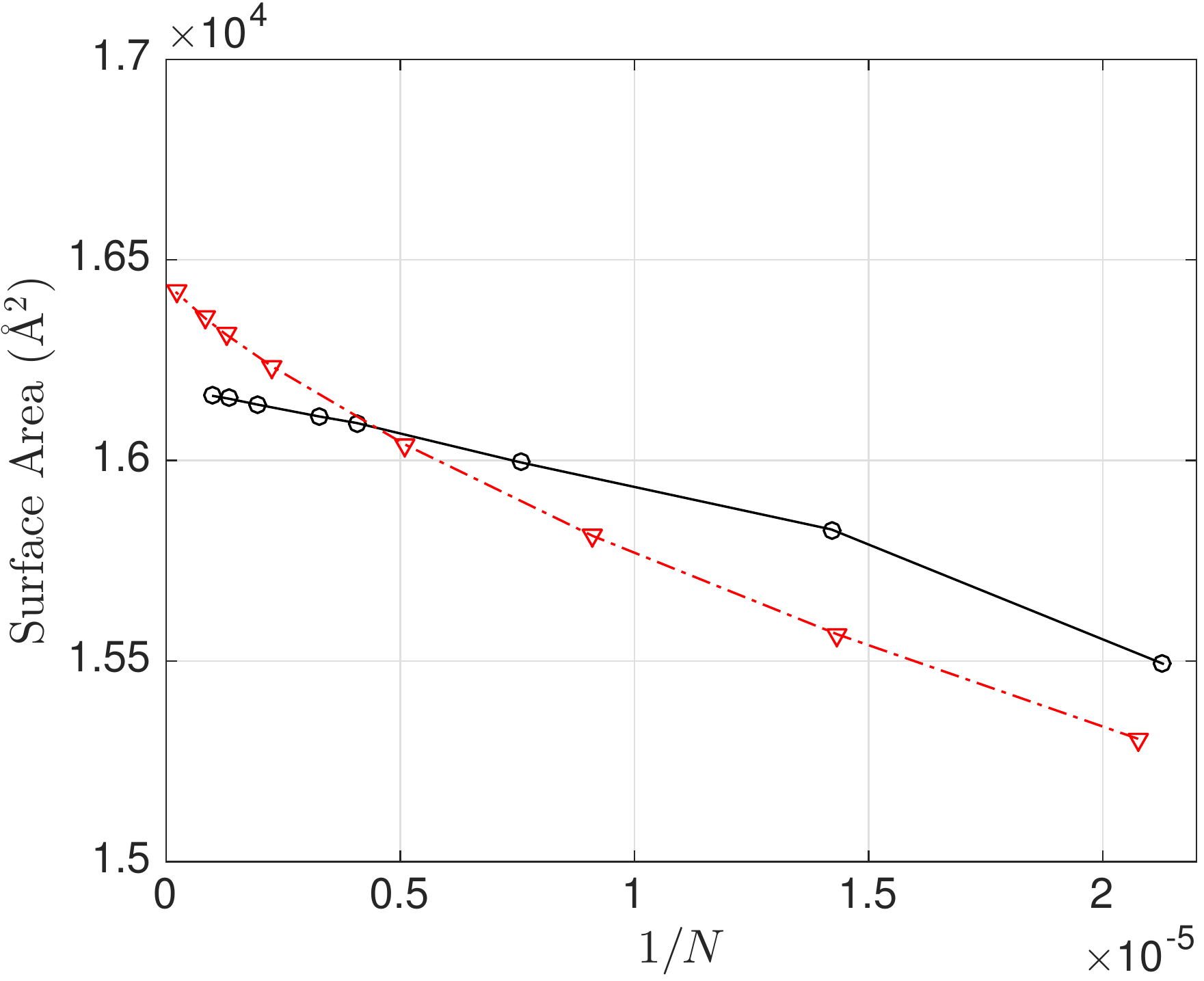}};
    \draw (-2, 2.5) node {(c) 3FR0};
  \end{tikzpicture} &
  \begin{tikzpicture}
    \draw (0, 0) node[inner sep=0] {\includegraphics[width=0.5\textwidth]{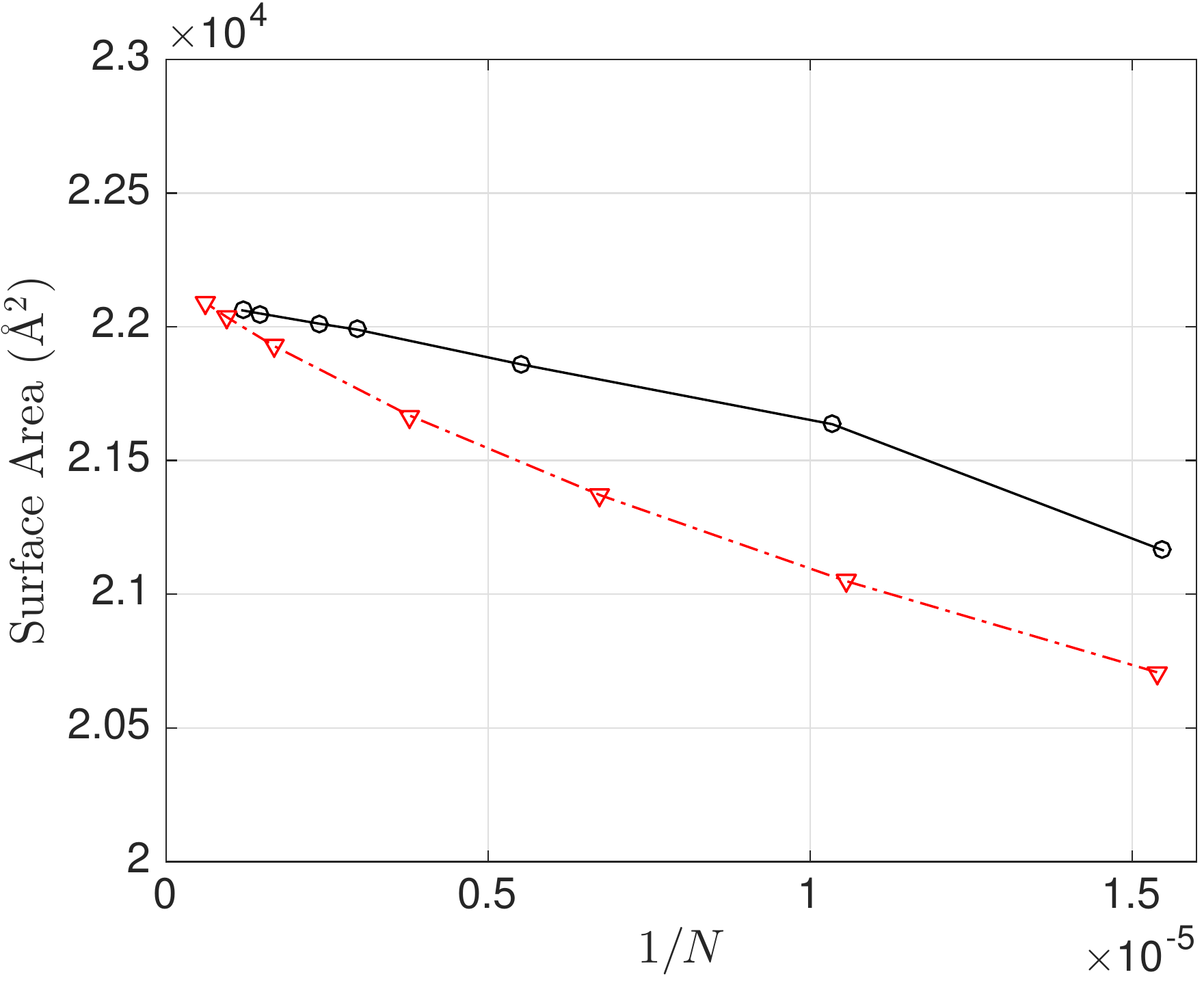}};
    \draw (-2, 2.5) node {(d) 1IL5};
  \end{tikzpicture} \\
\end{tabular}
\caption{
Surface area $S_a$ versus $N^{-1}$ for four representative proteins,
where $N$ is the number of triangles,
MSMS (black, solid line, $\circ$),
NanoShaper (red, dashed line, $\triangledown$).}
\label{fig:conv4}
\end{figure}

\begin{figure}[htb]
\centering
\begin{tabular}{cc}
  \begin{tikzpicture}
    \draw (0, 0) node[inner sep=0] {\includegraphics[width=0.5\textwidth]{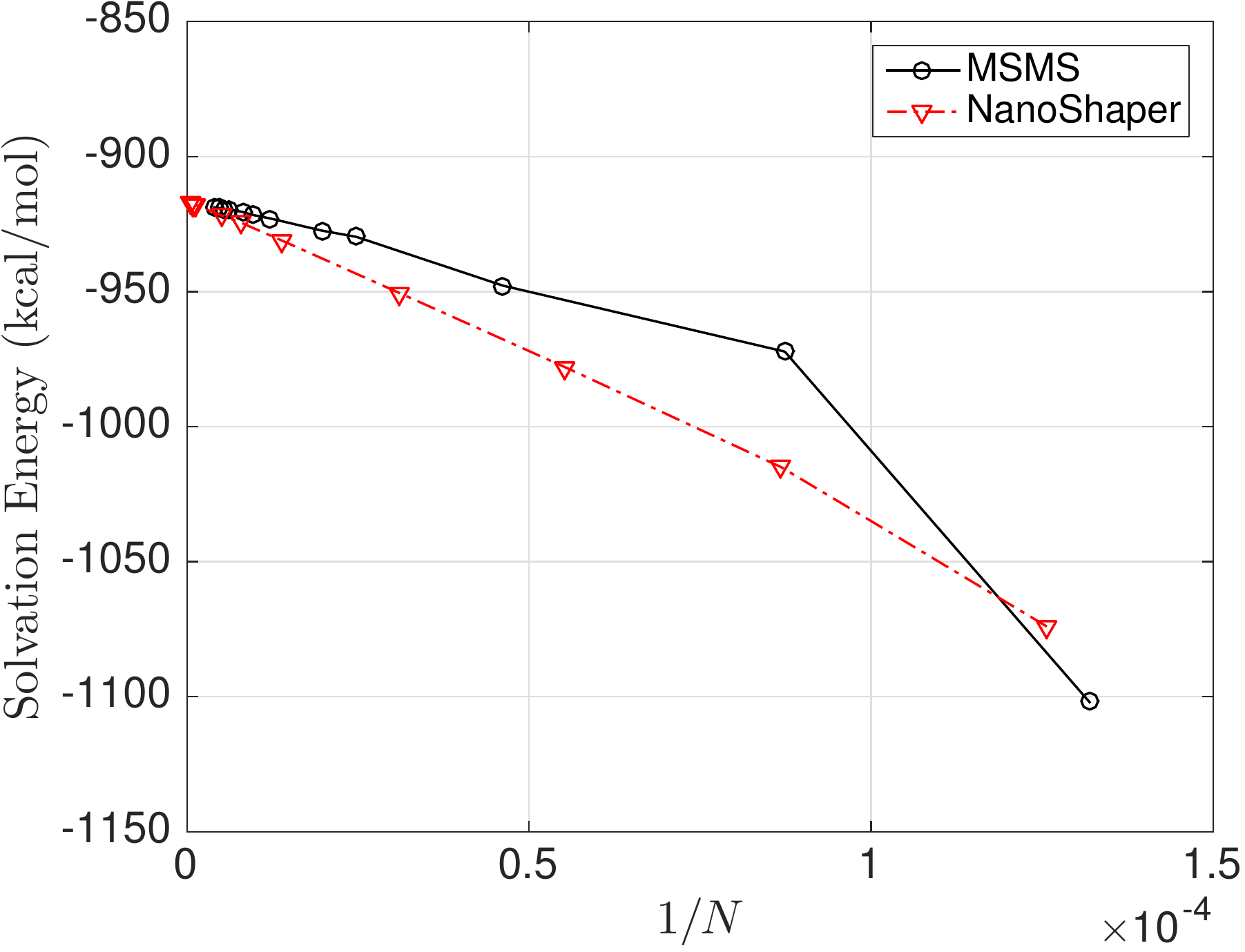}};
    \draw (-1.9, 2.5) node {(a) 1AIE};
  \end{tikzpicture} &
  \begin{tikzpicture}
    \draw (0, 0) node[inner sep=0] {\includegraphics[width=0.5\textwidth]{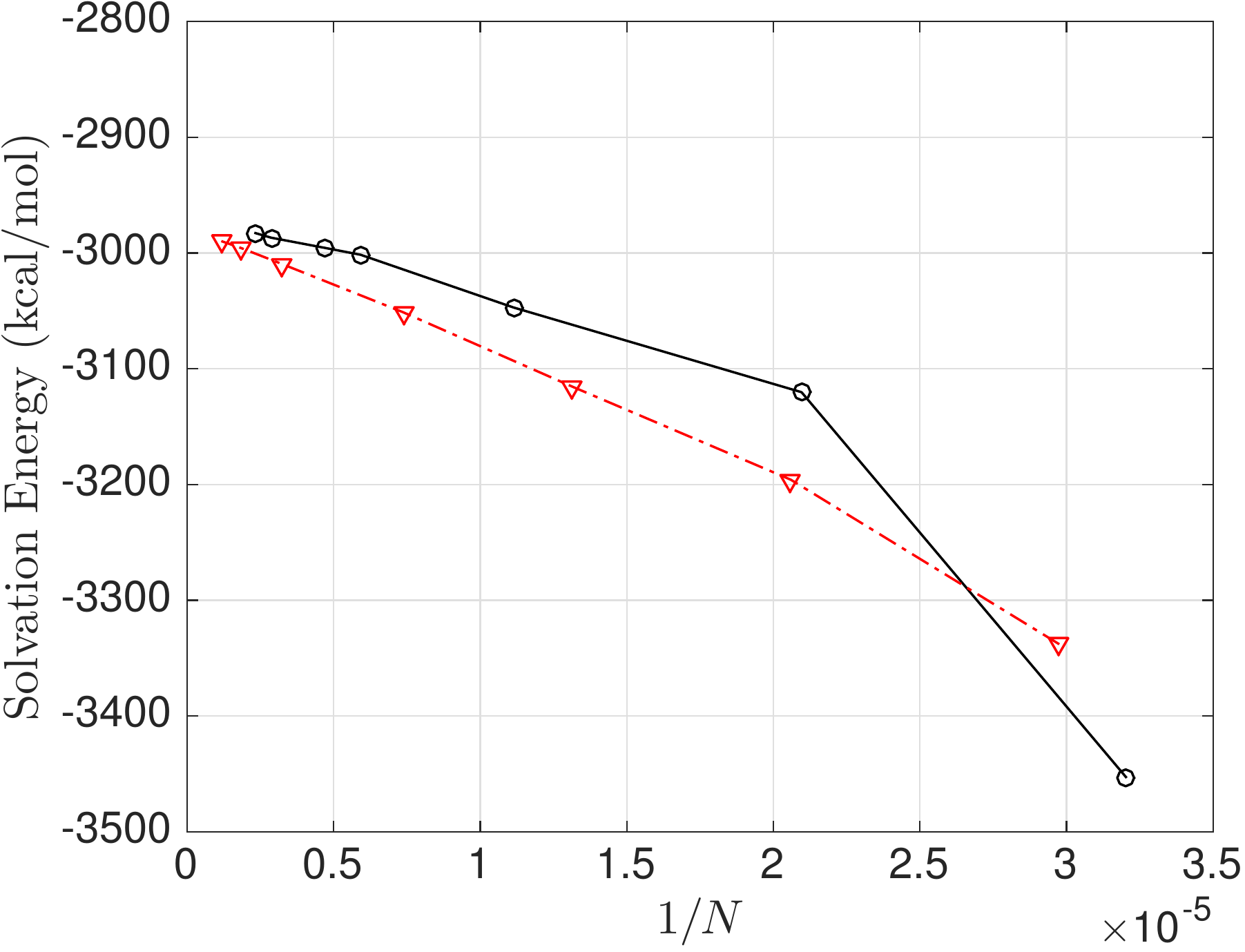}};
    \draw (-1.9, 2.5) node {(b) 1HG8};
  \end{tikzpicture} \\
  \begin{tikzpicture}
    \draw (0, 0) node[inner sep=0] {\includegraphics[width=0.5\textwidth]{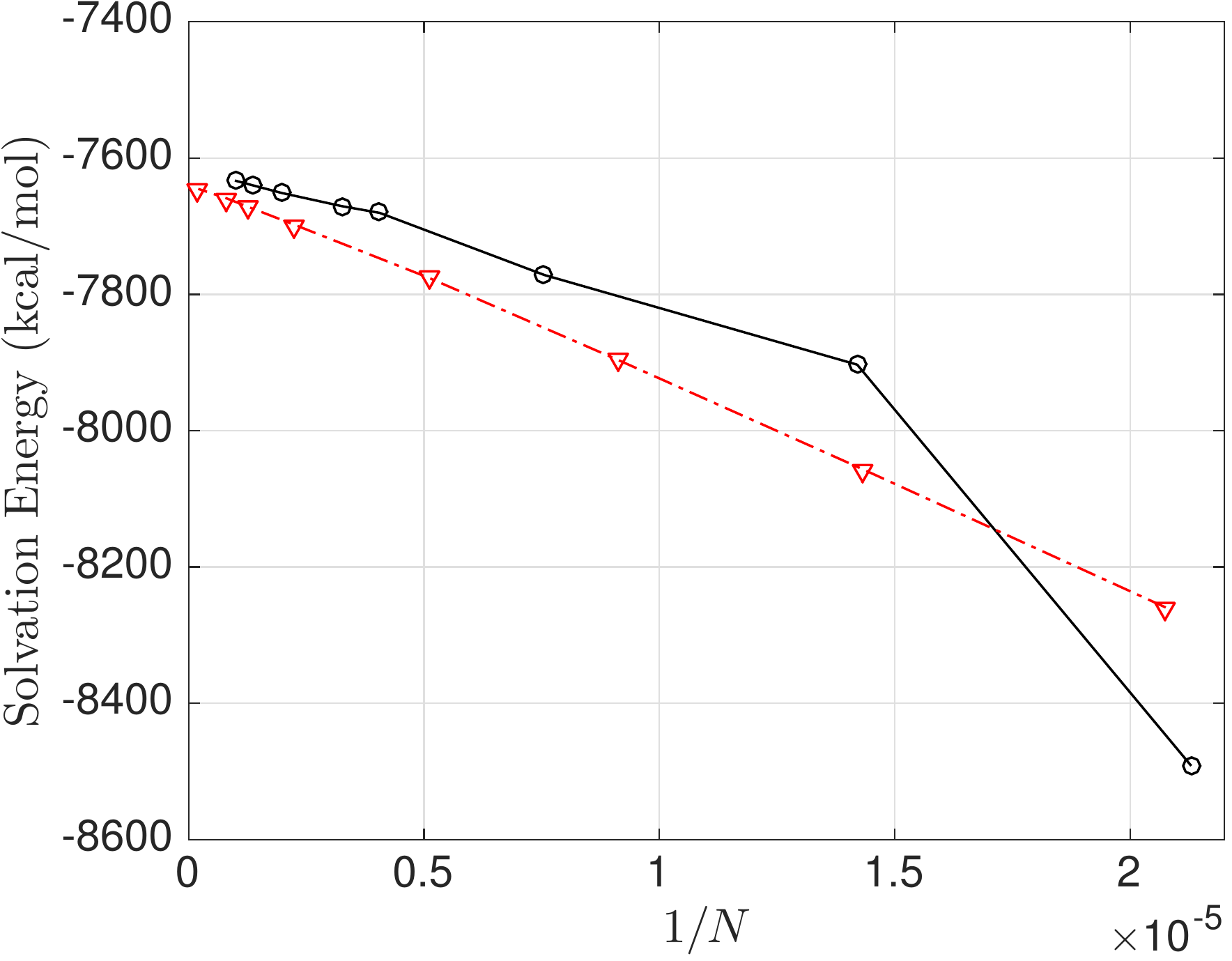}};
    \draw (-1.9, 2.5) node {(c) 3FR0};
  \end{tikzpicture} &
  \begin{tikzpicture}
    \draw (0, 0) node[inner sep=0] {\includegraphics[width=0.5\textwidth]{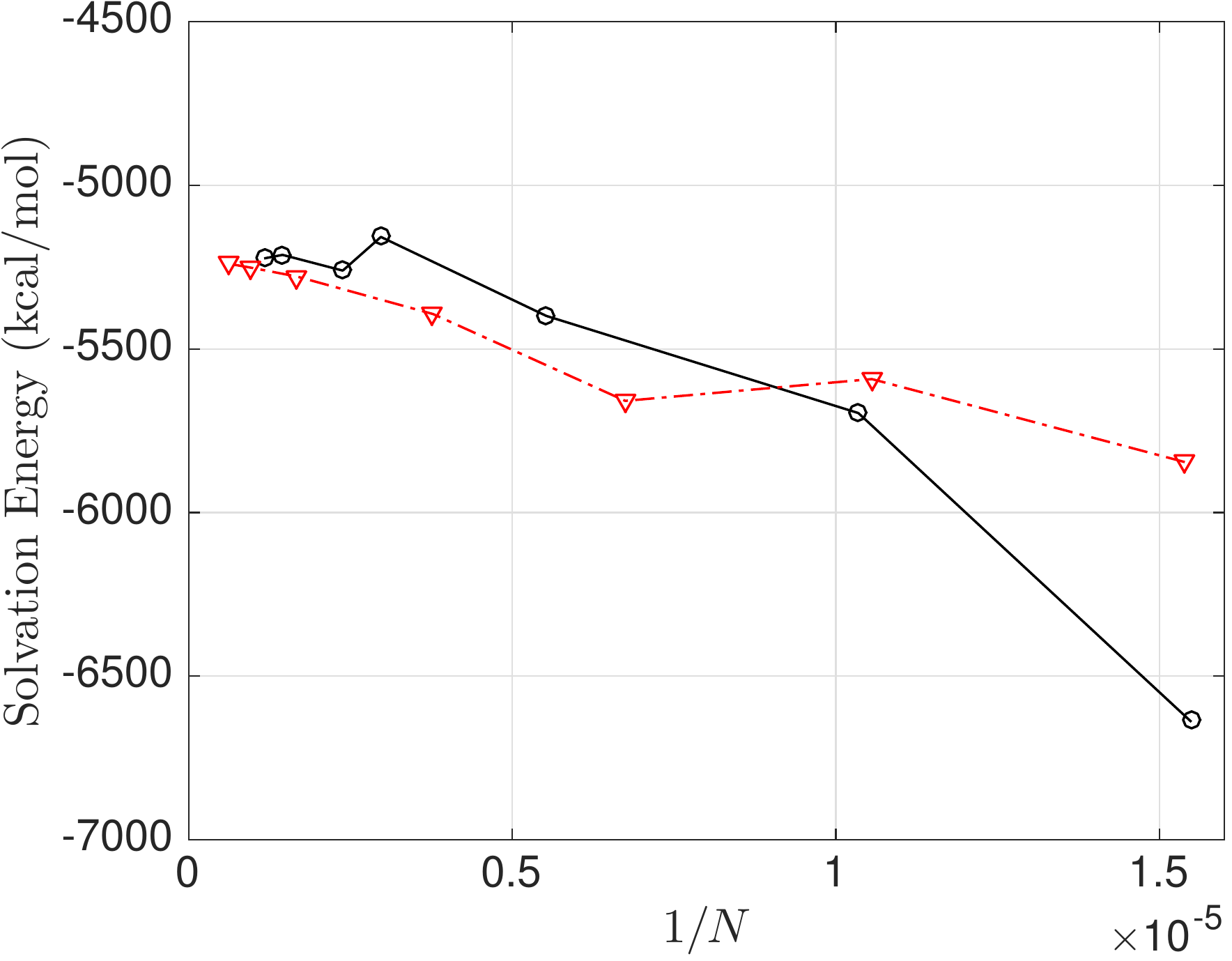}};
    \draw (-1.9, 2.5) node {(d) 1IL5};
  \end{tikzpicture} \\
\end{tabular}
\caption{
Solvation energy $E_{sol}$ versus $N^{-1}$ for four representative proteins, 
where $N$ is the number of triangles,
MSMS (black, solid line, $\circ$),
NanoShaper (red, dashed line, $\triangledown$).}
\label{fig:conv3}
\end{figure}

In this section we examine the dependence of the surface area $S_a$
and 
solvation energy $E_{\rm sol}$ on the mesh resolution 
for four representative proteins (1AIE, 1HG8, 3FR0, 1IL5).
Figure~\ref{fig:conv4} plots $S_a$ and Fig.~\ref{fig:conv3} plots $E_{\rm sol}$
versus $N^{-1}$, where $N$ is the number of surface elements.
As expected the MSMS and NanoShaper results for $S_a$ and $E_{\rm sol}$
converge to similar limits since both codes approximate the solvent excluded surface,
but several differences can be seen in their dependence on $N$.

First, concerning the surface area in Fig.~\ref{fig:conv4},
the MSMS and NanoShaper results converge to almost the same values in 
Fig.~\ref{fig:conv4}a,d (1AIE, 1IL5),
but in Fig.~\ref{fig:conv4}b,c (1HG8, 3FR0), 
the NanoShaper surface area is 2-3\% larger than the MSMS surface area.
In all cases the convergence with $N^{-1}$ is smooth.
The MSMS results approach their limit somewhat faster,
although MSMS was unable to generate reliable meshes with larger values of $N$;
either it fails to produce a mesh,
or the generated mesh was poorly formed.
The largest value we obtained using MSMS was $N \approx 2$e$+6$,
whereas NanoShaper had no such limitation.
Hence if it is necessary to generate a very dense mesh, 
or even a less dense mesh for a biomolecule with a large surface area,
then NanoShaper has an advantage.

Second, concerning the solvation energy in Fig.~\ref{fig:conv3},
the MSMS and NanoShaper results converge to almost the same value.
The MSMS results again approach their limiting value somewhat faster than the NanoShaper results, 
although the NanoShaper dependence on $N$ is smoother than the MSMS dependence.

\begin{figure}[htb]
\centering
\begin{tabular}{cc}
\includegraphics[width=0.5\linewidth]{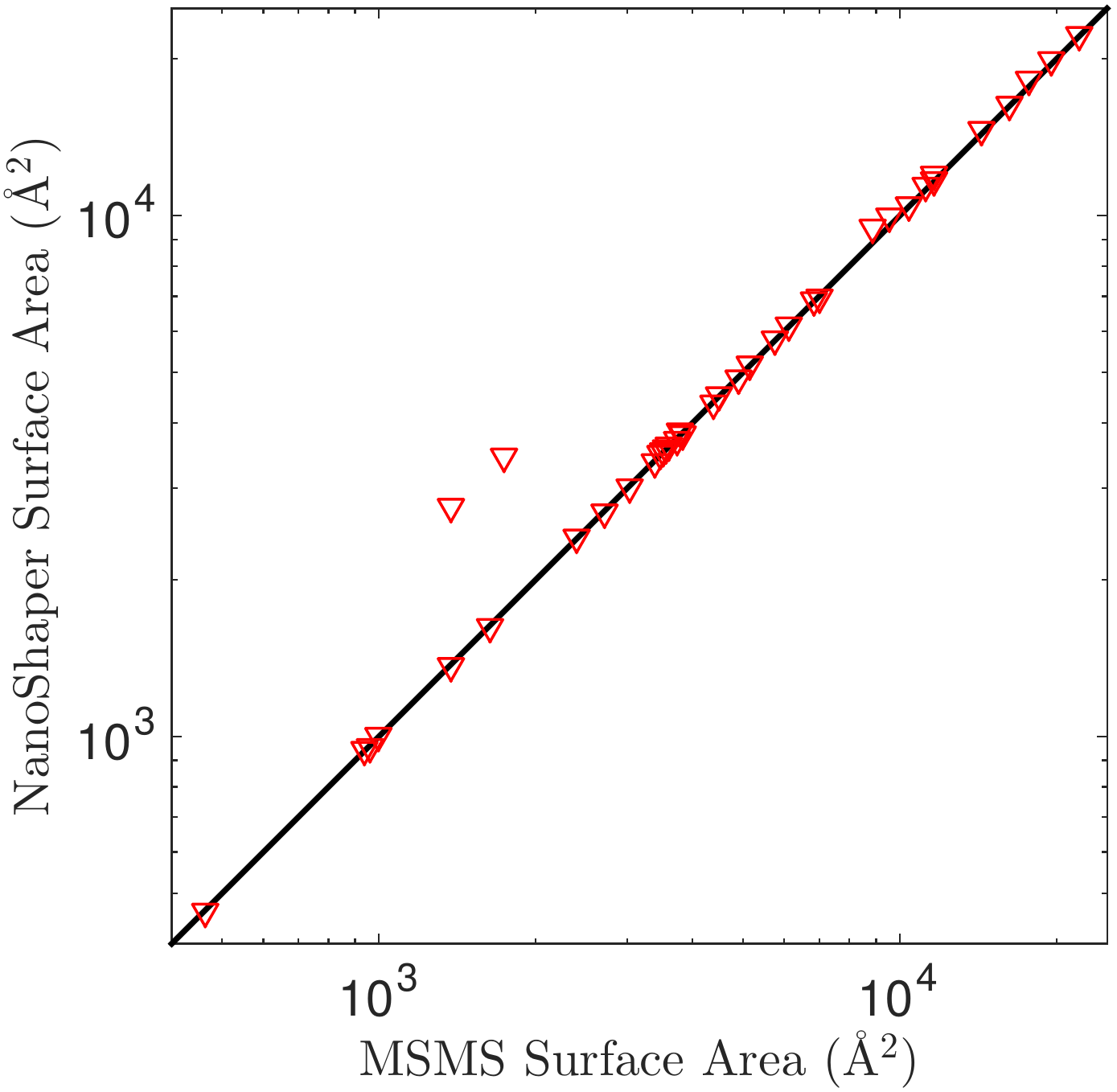} &
\includegraphics[width=0.5\linewidth]{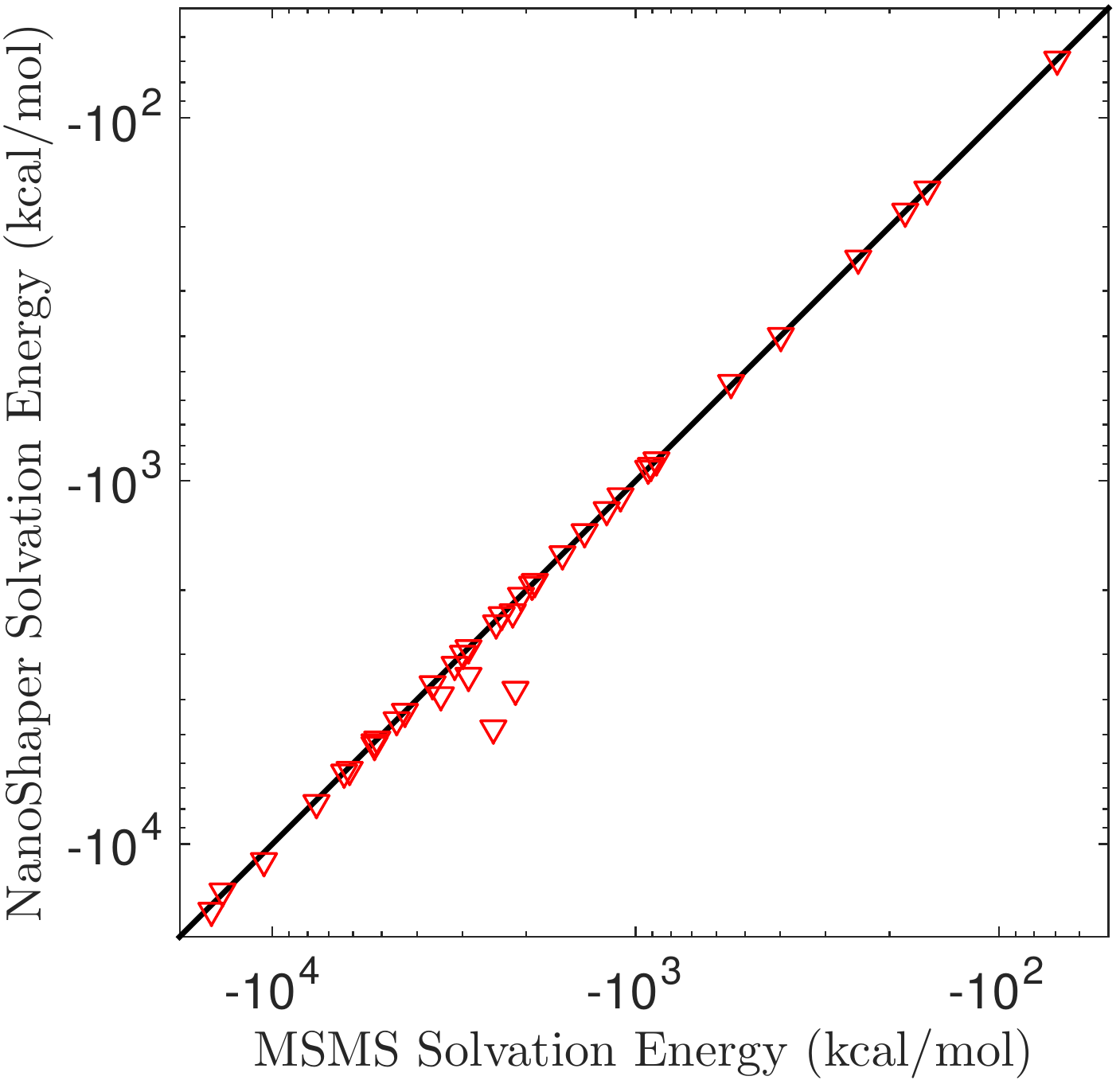} \\
(a) & (b) \\
\end{tabular}
\caption{
NanoShaper versus MSMS results for entire set of 38 biomolecules using
values extrapolated to the limit $N \to \infty$,
(a) surface area $S_a$,
(b) solvation energy $E_{\rm sol}$,
black lines indicate perfect correspondence.}
\label{fig:msmscomp1}
\end{figure}

Figure~\ref{fig:msmscomp1} 
displays the (a) surface area $S_a$ and (b) solvation energy $E_{\rm sol}$
for the entire set of 38 biomolecules,
where the NanoShaper results are plotted versus MSMS results.
In this case to reduce the dependence of the computed values on the mesh resolution $N$,
we extrapolated the computed $S_a$ and $E_{\rm sol}$ 
to the limit $N \to \infty$ using the two highest resolution meshes,
density $d=8, 16$ for MSMS
and
scaling factor $s = 4, 5$ for NanoShaper.
The correspondence between MSMS and NanoShaper results in Fig.~\ref{fig:msmscomp1}
is very good,
except for two molecules, 
1I2X and 375D, which consist of multiple domains, 
for which MSMS did not generate an accurate mesh.
These two anomalous cases are indicated by
the two markers furthest away from the diagonal line in Figs.~\ref{fig:msmscomp1}a,b.
In addition,
MSMS failed to produce surfaces for 13 other runs, 
and 
produced highly distorted surfaces with spurious solvation energy for 3 more runs.
These 16 spurious runs were removed from the calculations in this section.  
By contrast,
NanoShaper failed in only one case,
a low resolution mesh with scaling factor $s=1$
for the smallest molecule in the test set (2LWC, 75 atoms). 

\section*{\sffamily \Large Computational Efficiency}

\begin{figure}[htb]
\centering
\begin{tabular}{cc}
\includegraphics[width=0.5\linewidth]{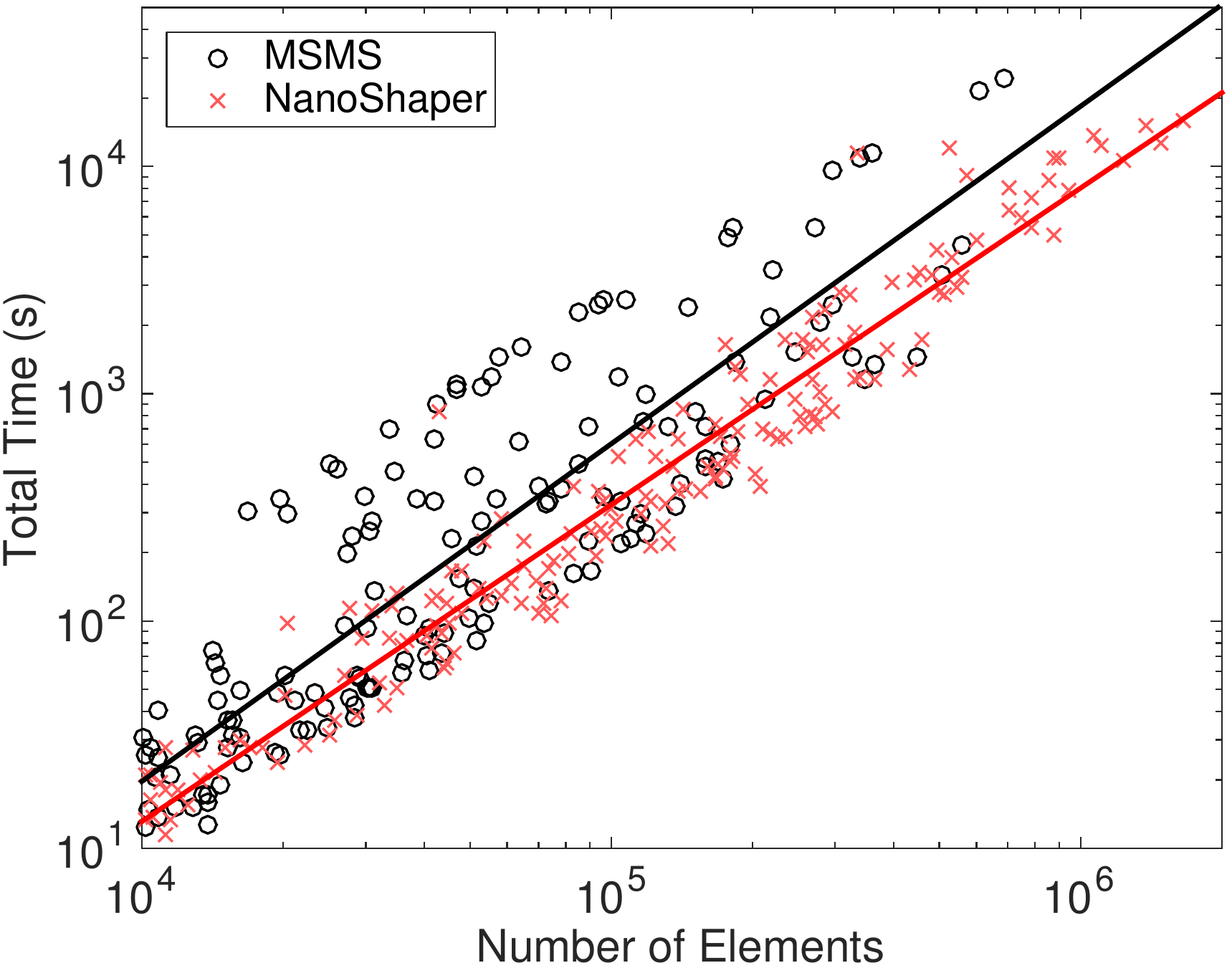} &
\includegraphics[width=0.5\linewidth]{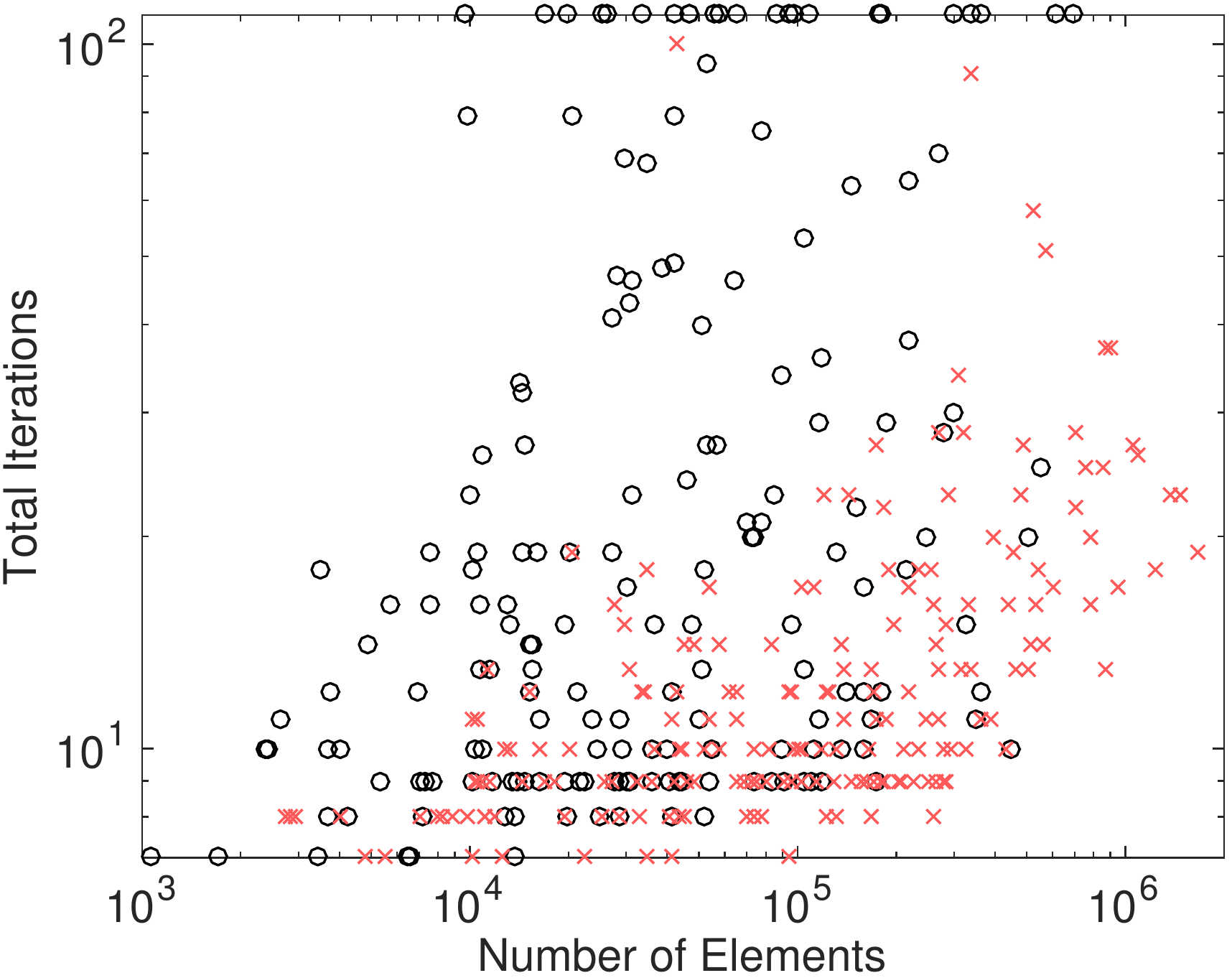}  \\
(a) & (b) \\
\end{tabular}
\caption{
Computational efficiency,
total run time of TABI-PB for computing solvation energy $E_{sol}$
using MSMS ($\circ$, black) and NanoShaper ($\times$, red) versus number of triangles $N$,
(a) run time (s), 
solid lines are least squares fits,
(b) number of GMRES iterations (maximum 110).}
\label{fig:timecomp2}
\end{figure}

Figure \ref{fig:timecomp2}a displays the total TABI-PB run time versus
the number of triangles $N$ for computing the solvation energy $E_{sol}$
using MSMS and NanoShaper meshes,
where the solid lines are least squares fits to the data.
The run time for creating and filtering the meshes 
is less than eight seconds in all cases, 
and thus constitutes a negligible fraction of the total run time.
The results show that in general,
NanoShaper meshes require less run time than MSMS meshes. 
This is supported by Fig.~\ref{fig:timecomp2}b 
showing the number of GMRES iterations in each case,
where the maximum number of iterations was set to~110.
The results show that in general,
NanoShaper meshes require fewer GMRES iterations than MSMS meshes.
Moreover, in the case of MSMS, the iteration limit was reached in 23 out 177 meshes,
while in the case of NanoShaper, the iteration limit was never reached.

Table~\ref{table:avgiter} displays
the average run time 
and
average number of GMRES iterations 
per triangle for each mesh type over the entire set of 38 biomolecules.
The results show that NanoShaper meshes require about 2/3 of the run time
and
1/4 of the number of iterations required by MSMS meshes.
The larger number of GMRES iterations required for MSMS meshes 
is attributed to the presence of triangles with large aspect ratio,
as seen in Fig.~\ref{fig:aspectratio},
and
irregular features in the generated surfaces,
as seen in Fig.~\ref{fig:surfacecomp2}.

\begin{table}[htb]
\caption{Average run time (s) and average number of GMRES iterations per triangle
for MSMS and NanoShaper meshes over entire set of 38 biomolecules.} 
\label{table:avgiter}
\begin{center}
\begin{tabular}{c|c|c}
\hline
 & average run time (s)/triangle & average iterations/triangle \\
\hline
MSMS 		& 6.67e--3		 & 1.17e--3	\\
NanoShaper	& 4.19e--3		 & 2.92e--4  \\
\hline
\end{tabular}
\end{center} 
\end{table}

\section*{\sffamily \Large CONCLUSIONS}

We compared the performance of MSMS and NanoShaper,
two widely used codes for
triangulating the solvent excluded surface (SES) 
in Poisson--Boltzmann simulations of solvated biomolecules.
Comparisons were made of the surface area
and
electrostatic solvation energy,
where the latter calculations were performed using the 
treecode-accelerated boundary integral (TABI-PB) solver
which utilizes a well-conditioned boundary integral formulation 
and 
centroid collocation on the SES triangulation.
In these calculations,
the linear system for the electrostatic potential and its normal derivative
on the SES is solved by GMRES iteration.
The matrix-vector product in each step of GMRES is computed by a treecode
which reduces the computational cost from $O(N^2)$ to $O(N\log N)$, 
where $N$ is the number of elements in the surface triangulation.

The MSMS and NanoShaper codes were compared for a test set of 38 biomolecules.
The meshes produced by the two codes are qualitatively similar,
although the MSMS meshes often contained triangles of exceedingly small area 
and high aspect ratio.
The computed values of the surface area and solvation energy produced by
MSMS and NanoShaper meshes often agree to within several percent. 
NanoShaper meshes were more computationally efficient, 
requiring less run time and fewer GMRES iterations than MSMS meshes. 
Furthermore, 
NanoShaper was consistently able to produce higher resolution meshes than MSMS,
and 
NanoShaper solvation energies exhibited smoother convergence with increasing mesh resolution.
A version of TABI-PB using NanoShaper was recently implemented 
as an option in APBS~\cite{Jurrus:2017aa}.

\section*{\sffamily \Large ACKNOWLEDGMENTS}

This work was supported by NSF grant DMS-1819094
and
a catalyst grant from the Michigan Institute for Computational Discovery and Engineering (MICDE).
Leighton Wilson was supported by the Department of Defense (DoD) through the 
National Defense Science \& Engineering Graduate Fellowship (NDSEG) Program.

\clearpage


\bibliography{library}



\end{document}